\documentclass[preprint]{elsarticle}

\usepackage[cmex10]{amsmath}
\usepackage{epsfig}
\usepackage{url}
\usepackage{psfrag}
\usepackage{amsfonts,amssymb}
\usepackage{graphicx}
\usepackage{framed}
\usepackage{epstopdf}
\usepackage{tikz}
\usepackage{array}
\usepackage{mathtools}
\usepackage{subfig}
\usepackage{steinmetz}
\usepackage{balance}
\usepackage{etex}
\usepackage{url}
\usepackage{color}

\usepackage[usenames,dvipsnames]{pstricks}
\usepackage{pst-grad} 
\usepackage{pst-plot} 

\journal{Chaos, Solitons \& Fractals}

\begin{document} 

\begin{frontmatter}

\title{Stability and non-linear dynamics of Dual congestion control schemes with two delays\tnoteref{t1}}
\tnotetext[t1]{Declarations of interest: none.}

\author[iitm]{Abuthahir\corref{mycorrespondingauthor}}
\ead{ee12d207@ee.iitm.ac.in}

 \author[iitm]{Gaurav Raina}
\ead{gaurav@ee.iitm.ac.in}

\address[iitm]{Department of Electrical Engineering, Indian Institute of Technology Madras,\\ Chennai-600036, India}
\cortext[mycorrespondingauthor]{Corresponding author at: Department of Electrical Engineering, Indian Institute of Technology Madras, Chennai-600036, India. }

\begin{abstract}
In this paper, we analyze some local stability and local bifurcation properties of the Proportionally fair, TCP fair, and the Delay-based dual algorithms in the presence of two distinct time delays. In particular, our focus is on the interplay between different notions of fairness, stability, and bifurcation theoretic properties.  Different notions of fairness give rise to different non-linear models for the class of Dual algorithms. One can devise conditions for local stability, for each of these models, but such conditions do not offer clear design recommendations on which fairness criteria is desirable. With a bifurcation-theoretic analysis, we have to take non-linear terms into consideration, which helps to learn additional dynamical properties of the various systems. In the case of TCP fair and Delay dual algorithms, with two delays, we present evidence that they can undergo a sub-critical Hopf bifurcation, which has not been previously revealed through analysis of the single delay variants of these algorithms. A sub-critical Hopf bifurcation can result in either large amplitude limit cycles or unstable limit cycles, and hence should be avoided in engineering applications. In the case of the Proportionally fair algorithm, we provide strong evidence to suggest that all one should expect is the occurrence of a super-critical Hopf bifurcation, which leads to stable limit cycles with small amplitude. Thus, from a design perspective, our analysis favors the use of Proportional fairness in the class of dual congestion control algorithms. To best of our knowledge, this is the first study that presents evidence to suggest that fluid models representing Internet congestion control algorithms may undergo a sub-critical Hopf bifurcation.
\end{abstract}


\end{frontmatter}




\section{Introduction}
Feedback plays a rather important role in the regulation of dynamical systems. If such feedback is time-delayed, then there are considerable implications for the stability and dynamical properties of the underlying system; for example, see \cite{conghuancsf2018,novitzkysiads2019,pender2018,shi2017} and references therein. Congestion control is an important component for the efficient operation of the Internet. Internet congestion control has been an active area of networking research for several decades; see \citep{srikant2012} for an overview. Congestion control protocols aim to control and regulate congestion by adjusting source rates based on feedback from the network. The presence of time-delays, in feedback related to congestion signals, has implications for both stability and bifurcation phenomena. There is a continued interest in analyzing the stability and dynamical properties of fluid models for congestion control algorithms \cite{ding2014,dongcsf2016,khoshcsf2019,raina2005}.    

In this paper, we focus on a particular class of congestion control algorithms, referred to as the Dual algorithms, where the congestion measure or price is determined by the resource and sent back to the source. These congestion signals prompt sources to adjust their data rates. The reader is directed to \citep{kelly2003} for an overview of dual congestion control algorithms. Two important design considerations in the dual algorithms are the fairness in resource allocation and system stability. The survey paper by \citep{kelly2003} provides an overview of the different notions of fairness. The fairness offered by the current transport protocol is often referred to as TCP fairness \citep{kelly2003}. Other competing notions are those of Proportional \citep{kelly1998} and Delay \citep{paganini2001} based fairness. In this paper, we consider three dual congestion control algorithms. These are the Fair dual with Proportional and TCP fairness as two cases, and the Delay dual. In the context of dual algorithms, an important design question to ask is which fairness criterion is advisable. We address this question using tools from control and bifurcation theory.

The feedback from the resource to end-points is time-delayed. Therefore, stability becomes a key concern when designing such algorithms. Previous studies of the stability of dual algorithms showed that the system undergoes a Hopf bifurcation, as the stability conditions are just violated. For example, see \citep{dingdelayhopfdual2009,ding2014,raina2005,tanghopfcontroldual2016ccc} for some stability analysis on integer-order models of dual algorithms, and \citep{huang2019,tangfracorder2017,xiaofractionalordertac2017, xiaofractionalorderautomatica2017} for some fractional-order counterparts. Also, one may ask: How sensitive are these systems to conditions of local stability? In the sense that if we perturbed them enough so that they just find themselves in a locally unstable regime, are some systems more sensitive to such a perturbation than others? An important design objective of the system is not only to ensure the local stability but also to make sure that any loss of stability always results in stable limit cycles of small amplitude. It is then advisable to choose algorithms, which produce stable limit cycles of small amplitude as it enters into the unstable region. A comprehensive understanding of local bifurcation phenomena may help yield insights into the dynamics of these algorithms in the unstable regime. Moreover, all the previous works  assumed that the delays experienced by all the flows are same. In this paper, we investigate the stability and Hopf bifurcation of the integer-order models of these algorithms with \textit{two} distinct delays.

In the stability analysis, we derive stability conditions that enable us to understand the role of various system parameters in ensuring stability. However, the stability results do not provide any design guidelines on which fairness is advisable. Thus, we proceed to analyze the consequences associated with the loss of stability. Using an exogenous non-dimensional bifurcation parameter, we show that the loss of stability occurs through a Hopf bifurcation \citep{hassard1981}. For the Proportionally fair dual algorithm, the analytical frameworks employed to investigate the nature of the Hopf bifurcation are the Poincar{\'e} normal form and the center manifold theorem \citep{hassard1981}, which are outlined in the Appendix.
We present strong evidence to highlight that the occurrence of a super-critical Hopf (stable limit cycles) is what we should expect as the local stability condition for the Proportionally fair algorithm gets violated.
Whereas for the TCP fair and Delay dual algorithms, unfortunately, the analytical solutions are complicated, and we resort to numerical techniques to analyze the nature of the Hopf bifurcation. The numerical analysis tool that we use is DDE-Biftool \citep{ddebiftool1,ddebiftool2}. The results of our numerical analysis show that the bifurcation in both the TCP fair and Delay dual algorithms can be either super-critical or sub-critical Hopf, depending on the parameter values. It is rather striking to find the possibility of a sub-critical Hopf bifurcation in dual algorithms. In fact, as far as we know, none of the fluid models of Internet congestion control have revealed so far the existence of a sub-critical Hopf bifurcation. A sub-critical Hopf would lead either to the emergence of unstable limit cycles, or to the emergence of limit cycles with very large amplitude \citep{strogatz2018}. Hence, the occurrence of a sub-critical Hopf bifurcation may lead to undesirable system behavior in engineering applications. Therefore, our results suggest that it is preferable to go with the Proportionally fair controller for dual algorithms.

In \citep{raina2005}, it was shown that the type of Hopf bifurcation for the single delay models of all these algorithms is \textit{always} super-critical. However, this paper considers the model with two distinct delays, and presents evidence for the possibility of a sub-critical Hopf bifurcation in the case of TCP fair and Delay dual algorithms. Therefore, the use case of delay heterogeneity considered in this paper reveals some novel insights into the dynamical properties of these algorithms. In addition, in a scalar non-linear equation with two discrete delays (see Example (A.1) in the Appendix), we showed analytically that it may be possible to change the nature of the Hopf bifurcation by simply changing the values of the delays.

The rest of this paper is organized as follows. In Section 2, we outline the models for the Fair and the Delay-based dual algorithms, and recapitulate some previous results related to a single delay system. In Section 3, we derive conditions to ensure local stability. In Section 4, we conduct a local Hopf bifurcation analysis, which enables us to characterize the type of bifurcation. In Section 5, we conclude with a summary of our contributions. For ease of exposition, the theoretical framework for the Hopf bifurcation analysis is contained in the Appendix.

\newcommand{\myk}{\tilde{a}}
\section{Models}
At the fluid level, casting the congestion control algorithms within the framework of delay differential equations has enabled their analysis to be subjected to tools from control theory with much success. For example, see \citep{huang2019,kelly2003,novitzkysiads2019,paganini2001,raina2005}. 

We now describe the integer-order fluid models of the Proportionally fair, TCP fair, and the Delay-based dual algorithms. 
\subsection{Single Delay Models}
Let us consider the following representation of the dual algorithms \citep{raina2005}
{\begin{align}
 \frac{d}{dt}p(t) &= \kappa p(t)^m \Big(x(t-\tau) - CI_{[p(t)>0]}\Big), \label{eq:1dualMainmodel}
\end{align}}\hspace{-5pt}
where $p \ge 0$ denotes the link price, $C$ is the link capacity, $\kappa > 0$ is the gain, $x(t) = \mathcal{D}\big( p(t) \big)$, $\tau$ denotes the round-trip time (RTT) which is the sum of all the delays from source to link, and from link to source. Here $\mathcal{D}(p)$ represents user demand function, and is non-negative and strictly decreasing. The demand function and the value of the parameter $m$ decide the fairness.
\subsubsection{Fair Dual:}
The Fair dual algorithms correspond to
\begin{align}
 m=1, \quad \mathcal{D}(p) = \left( \frac{w}{p} \right)^{1/\alpha}, \label{eq:1fairDualDemand}
\end{align}
which achieves $\alpha$-fair allocations \citep{kelly2003}. The parameter $w$ represents the user's willingness to pay. We will now describe how different parameter choices for $\alpha$ and $w$ give rise to Proportional fairness and TCP fairness.\\ \\
i) \textit{Proportional Fairness}\\
The parameter values of $\alpha=w=1$ yields Proportionally fair resource allocation \citep{kelly2003}. Thus, we obtain 
\begin{equation}
  x=\dfrac{1}{p}.
\end{equation}
Therefore the single delay model of Proportionally fair dual algorithm is given by
 \begin{equation}
 \frac{d}{dt}p(t) = \kappa p(t) \left(\frac{1}{p(t-\tau)} - C\right). \label{eq:1pfmodel} 
 \end{equation}
ii) \textit{TCP Fairness}\\
For $\alpha=2, w = 1/{\tau}^{2}$, we get a fairness provided by TCP \citep{kelly2003}, which yields  
\begin{equation}
  x=\dfrac{1}{\tau\sqrt{p}}.
\end{equation}
Thus, the model of TCP fair algorithm is given by
\begin{equation}
\frac{d}{dt}p(t) = \kappa p(t) \left(\frac{1}{\tau\sqrt{p(t-\tau)}}- C\right). \label{eq:1tcpmodel} 
\end{equation}
\subsubsection{Delay Dual:}
For the parameter value $m=0$, we obtain Delay dual, and the demand function is given by \citep{paganini2001}
\begin{align}
 \mathcal{D}(p) = D_{max}e^{-\alpha_{s}p/\tau} \label{eq:1expDemand}.
\end{align}
Here $D_{max}$ represents the maximum demand for the user, and the parameter $\alpha_{s}$ is chosen for stability. Then, the Delay dual algorithm can be represented as
\begin{equation}
 \frac{d}{dt}p(t) = \kappa \Big(D_{max}e^{-\alpha_{s}p(t-\tau)/\tau} - C\Big). \label{eq:1ddmodel}
\end{equation}

\begin{table*}[hbtp!]
\caption{Results for the Fair and the Delay dual algorithms with a single delay from \citep{raina2005}. Note that $\eta$, an exogenous parameter, was motivated to be the bifurcation parameter. The stability conditions associated with the different notions of fairness are not the same, but if satisfied local stability will be ensured. In the case of single delay, all the three algorithms always exhibit a super-critical Hopf bifurcation. However, they differ in the scaling factors of the amplitude of the emerging limit cycles..}\label{tab:1delay}

\newcommand{\fdtabfontsizeone}{\textbf}
\newcommand{\fdtabfontsizetwo}{\Large}
\centering
\resizebox{\columnwidth}{!}{%
\begingroup
\LARGE
\begin{tabular}{cccc}
\hline
\hline
\fdtabfontsizeone{Algorithm} & \fdtabfontsizeone{Sufficient} & \fdtabfontsizeone{Necessary \& sufficient } & \fdtabfontsizeone{Amplitude of the}\vspace{-1mm}\\  \fdtabfontsizeone{} & \fdtabfontsizeone{condition} & \fdtabfontsizeone{condition} & \fdtabfontsizeone{limit cycle}\\
\hline
\vspace{-3mm}
$ $ & $ $ & $ $ &  \\
\textbf{{Fair dual}}: {$x=\left(\dfrac{w}{p}\right)^{1/\alpha}$} $ \qquad $ &$ \dfrac{ \eta_c \kappa C \tau}{\alpha} < 1$ & $  \dfrac{\eta_c \kappa C \tau}{\alpha} < \dfrac{\pi}{2}$  & $\alpha p \sqrt{\dfrac{20\pi(\eta-\eta_c)}{3\pi-2}}$
\vspace{-1mm} 
\\ 
Proportional fairness: $\alpha=w=1$ & $ $ & $  $ & $ $ \vspace{-1mm} \\
TCP fairness: $\alpha=2,\ w= 1/\tau^2$ & $  $ & $  $ &   \vspace{-1mm}\\
\vspace{-1mm}
$ $ & $ $ & $ $ &  \\
\textbf{Delay dual}: $x= D_{max} e ^{{-p\alpha_s}/{\tau}}$ &$  \eta_c \kappa C \alpha_s < 1$ & $ \eta_c \kappa C \alpha_s < \dfrac{\pi}{2}$ & $ \dfrac{\tau}{\alpha_s} \sqrt{\dfrac{20\pi(\eta-\eta_c)}{3\pi-2}}$ \vspace{-1mm}\\
\\ 
\hline
\end{tabular}
\endgroup
}
\\
\end{table*}

Let us recall the key results when we have a single discrete delay \citep{raina2005}. The different demand functions and the impact they have on the amplitude of the bifurcating limit cycles is highlighted (from \citep{raina2005}) in Table \ref{tab:1delay}. Using Result III in the Appendix, one can derive some sufficient conditions for stability which has also been tabulated in Table \ref{tab:1delay}. Even the local bifurcation analysis of a scalar non-linear equation can highlight differences between competing algorithms.
In the case of single delay, both the Fair dual and Delay dual algorithms always exhibit a super-critical Hopf bifurcation, and lead to stable limit cycles. However, the Fair and the Delay dual algorithms have different scaling factors for the amplitude of their bifurcating limit cycles.
For the Fair dual algorithms, we find that the oscillations are proportional to the variable price. However, it was intriguing to discover the relationship of $\alpha$, which dictates fairness, to the size of the resulting limit cycles. Observe in Table \ref{tab:1delay}, that to leading order, the limit cycle of a TCP controller will have twice the amplitude of a Proportionally fair controller. For the delay dual, the bifurcating limit cycle is not proportional to the price. It is, in fact, proportional to the factor $\tau/\alpha_s$, where $\alpha_s$ is a control parameter whose values are chosen to ensure stability. 

Communication between geographically dispersed end-systems in large scale networks may have to contend with an arbitrary number of time-delayed feedback loops. As a step towards understanding the impact of delay heterogeneity on the performance of dual algorithms, we proceed to analyze the two delay variants of these algorithms.
\subsection{Two Delay Models}
We consider a single bottleneck topology where two users, located at different geographical locations, compete for a share of the scarce resource. The spatial diversity ensures that the users have different propagation delays. See Fig. \ref{fig:toydiagram1} for a pictorial representation of the model with a single bottleneck link and two different round-trip times. The two delay variants of the dual algorithms can be represented as 
{\begin{align}
 \frac{d}{dt}p(t) &= \kappa p(t)^m \Big(x(t-\tau_1)+ x(t-\tau_2) - CI_{[p(t)>0]}\Big), \label{eq:dualMainmodel}
\end{align}}\hspace{-5pt}
where $\tau_1$, $\tau_2$ are the round-trip times. Throughout this paper, we use the terms round-trip time and delay interchangeably.

\subsubsection{Proportional fairness:}
A system where we have two users with different round-trip times, each employing a Proportionally fair controller, yields
\begin{equation}
\frac{d}{dt}p(t) = \kappa p(t) \left(\frac{1}{p(t-\tau_1)}+ \frac{1}{p(t-\tau_2)} - C\right). \label{eq:pfmodel} 
\end{equation}
\subsubsection{TCP fairness:}
The model of the TCP fair algorithm with two delays is given by
\begin{equation}
\frac{d}{dt}p(t) = \kappa p(t) \left(\frac{1}{\tau_1\sqrt{p(t-\tau_1)}}+ \frac{1}{\tau_2\sqrt{p(t-\tau_2)}} - C\right). \label{eq:tcpmodel} 
\end{equation}
\subsubsection{Delay dual:}
Let us consider $D_{max,1}=D_{max,2}=D_{max}$ and $\alpha_{s,1}=\alpha_{s,2}=\alpha_{s}$. Then, the two delay model of the Delay dual algorithm is
\begin{equation}
 \frac{d}{dt}p(t) = \kappa \Big(D_{max}e^{-\alpha_{s}p(t-\tau_1)/\tau_1}+ D_{max}e^{-\alpha_{s}p(t-\tau_2)/\tau_2} - C\Big). \label{eq:ddmodel}
\end{equation}

\newcommand{\myarrowsize}{0.09cm 5.0}
\begin{figure*}[hbtp!]
\vspace{2mm}
\centering
\scalebox{0.6} 
{\hspace{-15mm}
\begin{pspicture}(0,-2.54083)(18.224112,2.54083)
\includegraphics[viewport= -129 15 -109 60, scale = 1.5]{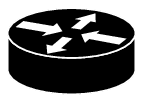}
\includegraphics[viewport= -225 15 -205 60, scale = 1.5]{router}

\psframe[linewidth=0.028222222,dimen=outer](3.015889,1.4791701)(2.544111,1.0073924)
\psline[linewidth=0.028222222cm,arrowsize=\myarrowsize,arrowlength=1.4,arrowinset=0.4]{->}(3.05,1.2232813)(4.653681,0.137328124)
\psline[linewidth=0.0628222222cm](6.5,-0.047)(10.841,-0.047) 

\psframe[linewidth=0.028222222,dimen=outer](3.015889,-0.98082983)(2.544111,-1.4526076)
\psline[linewidth=0.028222222cm,arrowsize=\myarrowsize,arrowlength=1.4,arrowinset=0.4]{->}(3.05,-1.1967187)(4.653681,-0.34671876)
\psline[linewidth=0.028222222cm,linestyle=dashed,dash=0.16cm 0.16cm](0.65,2.47)(17.61,2.47) 
\psline[linewidth=0.028222222cm,linestyle=dashed,dash=0.16cm 0.16cm](0.65,-2.6067189)(17.61,-2.6067189)  
\usefont{T1}{ptm}{m}{n}
\rput(2.32314062,0.67328125){\begin{Large}Source $1$ \end{Large} }
\usefont{T1}{ptm}{m}{n}
\rput(2.32314062,-0.7267187){\begin{Large}Source $2$  \end{Large}}

\usefont{T1}{ptm}{m}{n}
\rput(2.35957812,1.75){\begin{Large}Demand, $\mathcal{D}_1{(p)}$ \end{Large}}
\usefont{T1}{ptm}{m}{n}
\rput(2.35957812,-1.90567187){\begin{Large}Demand, $\mathcal{D}_2{(p)}$ \end{Large}}

\usefont{T1}{ptm}{m}{n}
\rput(8.957812,3.0){\begin{Large}Round-trip time, $\tau_1$ \end{Large}}
\usefont{T1}{ptm}{m}{n}
\rput(8.737812,-3.1567187){\begin{Large} Round-trip time, $\tau_2$ \end{Large}}
\usefont{T1}{ptm}{m}{n}
\rput(5.516957812,1.05){\begin{Large}Router\end{Large}}
\rput(11.6957812,1.05){\begin{Large}Router\end{Large}}
\rput(8.6957812,0.35){\begin{Large}Bottleneck link \end{Large}}
\usefont{T1}{ptm}{m}{n}
\rput(8.737812,-0.41567187){\begin{Large} Capacity, $C$ \end{Large}}
\psarc[linewidth=0.028222222,linestyle=dashed,dash=0.16cm 0.16cm](0.65,1.8232813){0.65}{90.0}{270.0}
\psarc[linewidth=0.028222222,linestyle=dashed,dash=0.16cm 0.16cm](0.65,-1.9767188){0.65}{90.0}{270.0}
\psline[linewidth=0.028222222cm,linestyle=dashed,dash=0.16cm 0.16cm,arrowsize=\myarrowsize,arrowlength=1.4,arrowinset=0.4]{->}(0.65,-1.3267188)(2.505,-1.3267188) 
\psline[linewidth=0.028222222cm,linestyle=dashed,dash=0.16cm 0.16cm,arrowsize=\myarrowsize,arrowlength=1.4,arrowinset=0.4]{->}(0.65,1.1732812)(2.505,1.1732812) 
\usefont{T1}{ptm}{m}{n}
\rput{-180.0}(30.648222,-2.4334376){\psframe[linewidth=0.028222222,dimen=outer](15.54,-0.96082985)(15.068222,-1.4326077)}
\psline[linewidth=0.028222222cm,arrowsize=\myarrowsize,arrowlength=1.4,arrowinset=0.4]{<-}(15.034111,-1.2132813)(12.6115,-0.28671875)
\rput{-180.0}(30.648222,2.4465625){\psframe[linewidth=0.028222222,dimen=outer](15.54,1.4791701)(15.068222,1.0073924)}
\psline[linewidth=0.028222222cm,arrowsize=\myarrowsize,arrowlength=1.4,arrowinset=0.4]{<-}(15.0541115,1.2867187)(12.6,0.111328125)
\rput{-180.0}(34.92,3.7265625){\psarc[linewidth=0.028222222,linestyle=dashed,dash=0.16cm 0.16cm](17.46,1.90){0.65}{90.0}{270.0}}
\psline[linewidth=0.028222222cm,linestyle=dashed,dash=0.16cm 0.16cm,arrowsize=\myarrowsize,arrowlength=1.4,arrowinset=0.4]{<-}(17.5,1.15)(15.71,1.15) 
\psline[linewidth=0.028222222cm,linestyle=dashed,dash=0.16cm 0.16cm,arrowsize=\myarrowsize,arrowlength=1.4,arrowinset=0.4]{<-}(17.58,-1.3267188)(15.718,-1.3267188)  
\rput{-180.0}(35.12,-3.7134376){\psarc[linewidth=0.028222222,linestyle=dashed,dash=0.16cm 0.16cm](17.61,-1.7567188){0.65}{90.0}{270.0}}
\usefont{T1}{ptm}{m}{n}
\rput(15.528,0.571171876){\begin{Large}Destination $1$ \end{Large}}
\usefont{T1}{ptm}{m}{n}
\rput(15.528,-0.68882812){\begin{Large}Destination $2$  \end{Large}}
\end{pspicture} 
}\\
\vspace{8mm}
\caption{{Two distinct sets of source-destination pairs with different round-trip times, share a single bottleneck link of capacity $C$. The user demand functions of source $1$ and $2$ are denoted as $\mathcal{D}_1(p)$ and $\mathcal{D}_2(p)$ respectively. Here, $\tau_1$, $\tau_2$ are the round-trip times.}}
\label{fig:toydiagram1}
\end{figure*} 
We now proceed to analyze these models.
\section{Local stability analysis}
In this section, we analyze the local stability of the Fair dual and the Delay dual algorithms with two discrete feedback delays. Instability in the system could be induced by varying some of the system parameters. But, variations of some system parameters may affect the equilibrium values, which is undesirable. Moreover, to conduct a unified analysis of all these algorithms, it is preferable to use a common exogenous parameter as the bifurcation parameter. So we introduce an exogenous and non-dimensional bifurcation parameter, $\eta$, to drive the system just into the unstable regime. Let us consider the perturbation $u(t) = p(t) - p^*$, where $p^*$ is the non-trivial equilibrium of the corresponding system. 
\subsection{Proportional fairness}
For the Proportionally fair dual algorithm, we conduct the local stability analysis using linearization. Since we analyze only the local stability, the linearization would give sufficient information to deduce whether or not the system converges to equilibrium. To linearize the non-linear system, we write the Taylor series expansion about the equilibrium point, and include only the linear terms. The linearized model of the non-linear differential equation (\ref{eq:pfmodel}), about the equilibrium $p^*$ is
\begin{equation}
\label{eq:linear_equation_pf}
\frac{d}{dt}u(t) = -\eta\kappa\myk\big(u(t-\tau_1)+u(t-\tau_2)\big),
\end{equation}
where
\begin{equation}
 \quad \myk = \frac{1}{p^*}, \quad p^*=\frac{2}{C}. 
\end{equation}
Looking for exponential solutions, the characteristic equation of~(\ref{eq:linear_equation_pf}) is given by
\begin{equation}
\lambda + \eta\kappa\myk\left(e^{-\lambda\tau_1}+e^{-\lambda\tau_2}\right) = 0,
\label{eq:chareq}
\end{equation}
where $\eta$, $\kappa$, $\myk$, $\tau_1$, $\tau_2>0$. For the system to be stable, all the roots of the characteristic equation should lie in the left half of the complex plane. 
When the round-trip times are zero, we get $\lambda = -2\eta\kappa\myk < 0$ and hence the system is asymptotically stable. 
However, when $\tau_1$, $\tau_2 > 0$ the roots may cross the imaginary axis for some values of the system parameters, and hence stability of the system cannot be guaranteed. 
Therefore, the condition for the crossover defines the bounds on the system parameters to maintain stability.
To find the critical condition, where this crossover occurs, we substitute $\lambda = \pm i\omega$, $\omega>0$ in (\ref{eq:chareq}). 
Equating the real and imaginary parts, we obtain 
\begin{equation}
\eta\kappa\myk\big(\cos(\omega\tau_1)+\cos(\omega\tau_2)\big) = 0, \label{eq:real0}
\end{equation}
\begin{equation}
\eta\kappa\myk\big(\sin(\omega\tau_1)+\sin(\omega\tau_2)\big) = \omega. \label{eq:img0}
\end{equation}
Solving~(\ref{eq:real0}) and~(\ref{eq:img0}), we get
$$\omega(\tau_1+\tau_2)=(2n+1)\pi,\qquad n=0,1,2,\cdots.$$
We only treat the case $n=0$, which gives $\omega_0=\pi/(\tau_1+\tau_2)$.
We now use the following theorem, stated in \citep{stepan1989}, to get the stability condition.
\newtheorem{theorem}{Theorem} 
\begin{theorem}\label{sufficient theorem}
\normalfont{\citep{stepan1989}}: \textit{The trivial solution of the scalar delay differential equation
\begin{equation}
\label{eq:stepaneqn1} 
\dot{x}(t) +bx(t-\tau_1)+bx(t-\tau_2)= 0
\end{equation}
is exponentially asymptotically stable if and only if}
\begin{equation}
\label{eq:stepaneqn2} 
0 < b < \dfrac{\pi}{2(\tau_1+\tau_2)\cos\left(\dfrac{\pi(\tau_1-\tau_2)}{2(\tau_1+\tau_2)}\right)}.
\end{equation}
\end{theorem}
Comparing (\ref{eq:linear_equation_pf}) with (\ref{eq:stepaneqn1}), the necessary and sufficient condition for~(\ref{eq:linear_equation_pf}) to be locally asymptotically stable can be obtained as 
\begin{equation}
\label{eq:ns_pf} 
\eta\kappa \myk(\tau_1+\tau_2)\cos\left(\dfrac{\pi(\tau_1-\tau_2)}{2(\tau_1+\tau_2)}\right)<\dfrac{\pi}{2}.
\end{equation}
Substituting the value of $\myk$ in~(\ref{eq:ns_pf}), we obtain the critical value of the bifurcation parameter, at which the system loses local stability as
\begin{equation}
\label{eq:kappa_cval} 
\eta_c=\dfrac{\pi}{\kappa C(\tau_1+\tau_2)\cos\left(\dfrac{\omega_0(\tau_1-\tau_2)}{2}\right)}.
\end{equation}
To show that the system undergoes a Hopf bifurcation at $\eta_c$, we need to satisfy the following transversality condition of the Hopf spectrum \citep{hassard1981}\\
$$\mathbf{Re}\left(\dfrac{d\lambda}{d\eta}\right)_{\eta=\eta_c}\neq 0.$$\\
Therefore evaluating
\begin{equation*}
\left(\dfrac{d\lambda}{d\eta}\right)_{\eta=\eta_c} = \frac{-\kappa\myk\left(e^{-\lambda\tau_1}+e^{-\lambda\tau_2}\right)}{1-\eta_c\kappa\myk\left(\tau_1 e^{-\lambda\tau_1}+ \tau_2 e^{-\lambda\tau_2}\right)}
\end{equation*}
we obtain 
\begin{equation*}
 \mathbf{Re}\left(\dfrac{d\lambda}{d\eta}\right)_{\eta=\eta_c} = \frac{ \omega_0 \kappa \myk (\tau_1 \sin(\omega_0 \tau_1) + \tau_2 \sin(\omega_0 \tau_2))}{\Phi_1^2+\Phi_2^2} \  > \ 0,
\end{equation*}
where
\begin{align*}
 \Phi_1 &= 1 - \eta_c \myk \kappa \tau_1 \cos(\omega_0 \tau_1) - \eta_c \myk \kappa \tau_2 \cos(\omega_0 \tau_2),\\ 
 \Phi_2 &= \eta_c \myk \kappa \tau_1 \sin(\omega_0 \tau_1) + \eta_c \myk \kappa \tau_2 \sin(\omega_0 \tau_2),\\ \\
 \eta_c &= \dfrac{\pi}{\kappa C(\tau_1+\tau_2)\cos\left(\dfrac{\omega_0(\tau_1-\tau_2)}{2}\right)}.
\end{align*}
Therefore, the Proportionally fair dual algorithm undergoes a Hopf bifurcation, at $\eta = \eta_c$, with period $2\pi/\omega_0$ where  $\omega_0=\pi/(\tau_1+\tau_2)$.

\begin{figure}[h!]
\centering
\psfrag{tau1}{$\tau_1$}
\psfrag{tau2}{$\tau_2$}
\includegraphics[trim=0cm 0cm 0cm 1cm, clip=true,width=3.25in,height=2.25in]{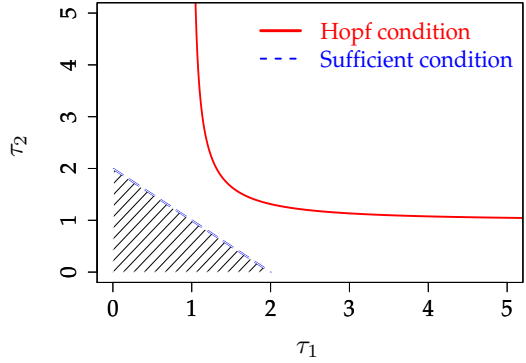}
\vspace{-2mm}
\caption{ Stability chart for the Proportionally fair dual algorithm, highlighting the Hopf condition and sufficient condition for local stability. The parameter values used are $\kappa = \eta = C = 1$.}
\label{fig:pfstbchart}
\end{figure}

We now state a simple sufficient condition for local stability, which can be obtained using Result II in the Appendix. Equation (\ref{eq:linear_equation_pf}) will be asymptotically stable if
\begin{equation}
\label{eq:s_pf} 
\frac{\eta \kappa C(\tau_1+\tau_2)}{2}<1.
\end{equation} 
Fig. \ref{fig:pfstbchart} shows the Hopf condition and the sufficient condition for local stability of the Proportionally fair dual algorithm.

\subsection{TCP fairness}
The non-trivial equilibrium of the TCP fair dual algorithms with two delays is given by 
\begin{equation}
 p^*=\left(\dfrac{1}{C\tau_1}+\dfrac{1}{C\tau_2}\right)^2.
 \label{eq:tcpeqbm}
\end{equation}
Linearizing (\ref{eq:tcpmodel}) about the equilibrium, we get
\begin{equation}
\label{eq:linear_equation_tcp}
\frac{d}{dt}u(t) = -\eta \kappa a_{1}u(t-\tau_1) - \eta \kappa a_{2}u(t-\tau_2),
\end{equation}
where
\begin{equation*}
 a_{1} = \frac{1}{2\tau_1\sqrt{p^*}}, \quad  a_{2} = \frac{1}{2\tau_2\sqrt{p^*}}.
\end{equation*}
The characteristic equation of~(\ref{eq:linear_equation_tcp}) is given by
\begin{equation}
\lambda + \eta \kappa a_{1}e^{-\lambda\tau_1}+ \eta \kappa a_{2}e^{-\lambda\tau_2} = 0,
\label{eq:chareq_tcp}
\end{equation}
where $\eta$, $\kappa$, $a_{1}$, $a_{2}$, $\tau_1$, $\tau_2>0$.
Substituting $\lambda = \pm i\omega$, $\omega>0$ in (\ref{eq:chareq_tcp}) and equating the real and imaginary parts to zero, we obtain 
\begin{equation}
\eta \kappa a_{1}\cos(\omega\tau_1)+ \eta \kappa a_{2}\cos(\omega\tau_2) = 0, \label{eq:real0_tcp}
\end{equation}
\begin{equation}
\eta \kappa a_{1}\sin(\omega\tau_1)+ \eta \kappa a_{2}\sin(\omega\tau_2) = \omega. \label{eq:img0_tcp}
\end{equation}

\begin{figure}[h!]
\centering
\psfrag{tau1}{$\tau_1$}
\psfrag{tau2}{$\tau_2$}
\includegraphics[trim=0cm 0cm 0cm 1cm, clip=true,width=3.25in,height=2.25in]{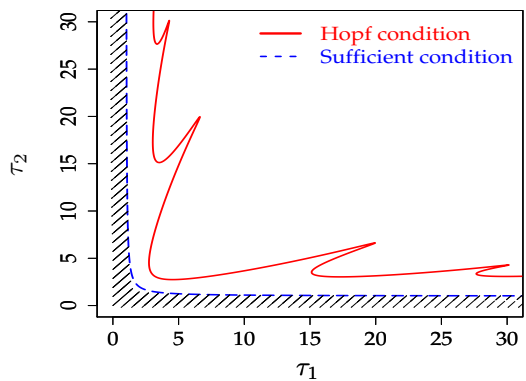}
\vspace{-2mm}
\caption{ Stability chart for the TCP fair dual algorithm, highlighting the Hopf condition and sufficient condition for local stability. The parameter values used are $\kappa = \eta = C = 1$.}
\label{fig:tcpstbchart}
\end{figure}

It is important to observe that the coefficients $a_1$ and $a_2$ are not equal and also delay-dependent.
In \citep{beretta1delay2002,xu1delaydep}, stability of the system with single delay and delay-dependent coefficients has been analyzed. In the two delay case, \citep{hale2delayindep,Piotrowska} analyze the differential equation with coefficients that are not equal and delay independent. But, in the two-delay case with delay dependent coefficients, there is no efficient method to analyze these equations. The complexity involved here is, when we remove $\tau_1$ to find the solutions, $\omega$ and $\tau_2$ are related by transcendental equation. So, if we fix $\tau_2$, and try to compute $\omega$, multiple solutions may occur, which makes the problem complex.
Therefore, in this case, it is difficult to find a closed-form expression for the necessary and sufficient condition. Numerical techniques allow insight into the cases where the analytical solutions are difficult to solve. So we should resort to some numerical tools that are able to trace the stability boundaries for the delay differential equations. We plot the Hopf boundary numerically using DDE-Biftool (a Matlab package for numerical stability and bifurcation analysis of delay differential equations).
Fig. \ref{fig:tcpstbchart} shows the Hopf condition in the parameter space of $\tau_1 $ and $\tau_2$, with $\kappa=\eta=C=1$.

We now try to derive a sufficient condition to ensure local stability. Using Result I in the Appendix, we can obtain the sufficient condition for \eqref{eq:linear_equation_tcp} to be stable as
\begin{equation}
\eta \kappa a_{1}\tau_1+ \eta \kappa a_{2}\tau_2 < 1.
\label{eq:tcpsuffcondn}
\end{equation}
For $\eta=1$, and substituting the values of $a_{1}$ and $a_{2}$ in \eqref{eq:tcpsuffcondn}, we get the sufficient condition for stability as 
\begin{equation}
 \frac{\kappa C \tau_1 \tau_2}{\tau_1 + \tau_2} < 1. \label{eq:suffcondtcp2}
\end{equation}
See  Fig. \ref{fig:tcpstbchart} for the graphical representation of the sufficient condition for the TCP fair dual algorithm

We can also obtain a common sufficient condition for the Proportional and the TCP fair controller, in terms of user demand functions. The linearized system of the fair dual algorithms with two delays is of the form  
\begin{equation} 
\label{eq:linear_equation_gen_fairdual}
\frac{d}{dt}u(t) = -\eta \kappa \frac{\mathcal{D}_1(p)}{\alpha}u(t-\tau_1) - \eta \kappa \frac{\mathcal{D}_2(p)}{\alpha}u(t-\tau_2),
\end{equation}
where $\mathcal{D}_r(p) = x_r = ({w_r}/{p})^{1/\alpha}$ is the demand function of user $r$.
We now state a sufficient condition for the local stability, which can be obtained by an analysis of \eqref{eq:linear_equation_gen_fairdual} using Result I in the Appendix.\\ \\
Equation \eqref{eq:linear_equation_gen_fairdual} is asymptotically stable if
\begin{equation}
 \eta \kappa C \overline{T}/\alpha < 1.
\end{equation}
where 
\begin{equation}
 \overline{T} = \frac{\sum_r{\mathcal{D}_r(p) \tau_r}}{\sum_r{\mathcal{D}_r(p)}} =\frac{\sum_r{x_r \tau_r}}{\sum_r{x_r}}
\end{equation}
is the average round-trip time of the packets through the resource.
\subsection{Delay dual}
The non-trivial equilibrium for (\ref{eq:ddmodel}) is given by 
\begin{equation}
\label{eq:eqbmdd}
e^{-\alpha_{s}p^*/\tau_1}+ e^{-\alpha_{s}p^*/\tau_2} = \frac{C}{D_{max}}.
\end{equation}
Linearizing (\ref{eq:ddmodel}) about the equilibrium, we get
\begin{equation}
\label{eq:linear_equation_dd}
\frac{d}{dt}u(t) = -\eta \kappa a_{1}u(t-\tau_1) - \eta \kappa a_{2}u(t-\tau_2),
\end{equation}
where
\begin{align}
 a_{1} &= \frac{\alpha_{s}D_{max}e^{-\alpha_{s}p^*/\tau_1}}{\tau_1} = \frac{\alpha_{s}D_1(p)}{\tau_1}\\ \nonumber
 a_{2} &= \frac{\alpha_{s}D_{max}e^{-\alpha_{s}p^*/\tau_2}}{\tau_2} = \frac{\alpha_{s}D_2(p)}{\tau_2}.
 \label{eq:a1a2dd}
\end{align}
We now plot the Hopf condition using DDE-Biftool (see Fig. \ref{fig:ddstbchart}). Using Result I in the Appendix, we can obtain the sufficient condition for stability as 
\begin{equation}
\label{eq:sufconddd}
\eta \kappa a_{1}\tau_1+ \eta \kappa a_{2}\tau_2 < 1.
\end{equation}
Substituting the values of $a_1$ and $a_2$ in \eqref{eq:sufconddd}, and then using \eqref{eq:eqbmdd}, we obtain the sufficient condition for local stability as
\begin{equation}
\label{eq:suffconddd1}
\kappa \alpha_s C <1.
\end{equation}
Note that the sufficient condition does not depend on the delays, as is the case for single delay. Fig. \ref{fig:ddstbchart} highlights the relationship between the parameter $\kappa$ and $\alpha_s$ in ensuring local stability, for the Delay dual algorithm.
\begin{figure}[h!]
\centering
\psfrag{c}{$\alpha_s$}
\psfrag{k}{$\kappa$}
\includegraphics[trim=0cm 0cm 0cm 1cm, clip=true,width=3.25in,height=2.25in]{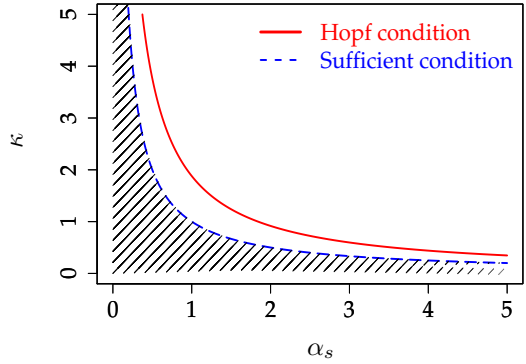}
\vspace{-2mm}
\caption{Stability chart for the Delay dual algorithm, highlighting the Hopf condition and sufficient condition for local stability. The values of the parameters chosen are $\eta = C = 1$.}
\label{fig:ddstbchart}
\end{figure}


\begin{table}[hbtp!]
\caption{Sufficient conditions to ensure local stability for the single delay and two delay dual algorithms. In contrary to the Fair dual algorithms, the sufficient condition for the Delay dual algorithm does not depend on the delays.}\label{tab:12delay}
\renewcommand{\arraystretch}{2.25}
\centering
\begin{tabular}{lll}
\hline
\hline
Algorithm   &  Single delay  & Two delays \\
\hline
\vspace{-10mm}\\
Proportional fairness & $\kappa C \tau < 1$ & $  \dfrac{\kappa C (\tau_1 + \tau_2)}{2} < 1$ \vspace{0mm} \\  
TCP fairness & $ \dfrac{\kappa C \tau}{2} < 1$ & $ \dfrac{\kappa C \tau_1\tau_2}{(\tau_1+\tau_2)} < 1$ \vspace{0mm}\\ 
Delay dual & $ \kappa C \alpha_s < 1$ & $ \kappa C \alpha_s < 1$ \\
\hline 
\hline
\end{tabular}
\\
\end{table}

As highlighted in the Table. \ref{tab:12delay}, the local stability conditions associated with the different notions of fairness are not the same, but if satisfied local stability will be ensured. However, based on these stability results we are unable to make any decision on which fairness is desirable. Moreover, any congestion control algorithm is not only to ensure local stability of the equilibrium, but also to make sure that any loss of stability, that may happen, results in \emph{stable} limit cycles of small amplitude. So it is natural to study the characteristics of the bifurcating periodic solutions. To that end, we conduct a local Hopf bifurcation analysis in the next section.
\section{Hopf bifurcation analysis}
In local stability analysis, we established some conditions for local stability for the system, coupled with the three different notions of fairness. In this section, we analyze the consequences associated with the loss of local stability. In particular, we are concerned with the loss of local stability occurring via a Hopf bifurcation leading to the onset of limit cycles, as the bifurcation parameter crosses a critical value.

We first note that the models that we consider are special cases of the following non-linear delay equation.
\begin{equation}
\label{eq:gen_non_eq0}
\dfrac{d}{dt}x(t) = \eta f\big(x(t),x(t-\tau_1),x(t-\tau_2)\big),
\end{equation}
where $f$ has a unique equilibrium denoted by $(x^*,y^*,z^*)$ and $\tau_1,\ \tau_2,\  \eta > 0$. Define $u(t) = x(t)-x^*,$ and take a Taylor series expansion for  (\ref{eq:gen_non_eq0}) including the linear, quadratic and cubic terms to obtain
\begin{equation}
\begin{aligned}
\dfrac{d}{dt}u(t)=\  &\eta \big(\xi_yu(t-\tau_1)+\xi_zu(t-\tau_2)+\xi_{xy}u(t)u(t-\tau_1)+\xi_{xz}u(t)u(t-\tau_2)\\
&+\xi_{yy}u^2(t-\tau_1)+\xi_{yz}u(t-\tau_1)u(t-\tau_2)+\xi_{zz}u^2(t-\tau_2)\\
&+\xi_{xyy}u(t)u^2(t-\tau_1)+\xi_{xzz}u(t)u^2(t-\tau_2)+\xi_{yyy}u^3(t-\tau_1)\\
&+\xi_{xyz}u(t)u(t-\tau_1)u(t-\tau_2)+\xi_{yyz}u^2(t-\tau_1)u(t-\tau_2)\\
&+\xi_{zzz}u^3(t-\tau_2)+\xi_{yzz}u(t-\tau_1)u^2(t-\tau_2)+ \mathcal{O}(u^4)\big)
\end{aligned}
\label{eq:linear_noneq0}
\end{equation}
where, letting $f^*$ denote evaluation of $f$ at $(x^*,y^*,z^*)$
\begin{alignat*}
\xi\xi_i&=f^*_i,&\xi_{ii}&=\dfrac{1}{2}f^*_{ii},&\xi_{iii}&=\dfrac{1}{6}f^*_{iii} &\forall \  i &\in \{x,y,z\} & \ \\
\xi_{xy}&=f^*_{xy},& \xi_{xz}&=f^*_{xz},& \xi_{yz}&=f^*_{yz},& \xi_{xxy}&=\dfrac{1}{2}f^*_{xxy},& \xi_{xyz}&=f^*_{xyz},\\
\xi_{xxz}&=\dfrac{1}{2}f^*_{xxz},\ \  & \xi_{xyy}&=\dfrac{1}{2}f^*_{xyy},\ \  &\xi_{xzz}&=\dfrac{1}{2}f^*_{xzz}, \ \  & \xi_{yyz}&=\dfrac{1}{2}f^*_{yyz},\ \  &\xi_{yzz}&=\dfrac{1}{2}f^*_{yzz}.
\end{alignat*}
\subsection{Proportional fairness}
We now consider the model of the Proportionally fair algorithm and perform the necessary
calculations to determine the type of Hopf bifurcation as local instability
just sets in. We opt for the method of Poincar{\'e} normal forms and the Center Manifold Theorem (see \citep{hassard1981} for details) to analyze the nature of the Hopf bifurcation. The requisite calculations, which are presented in the Appendix assume that the reader is well versed in the pertinent theory. For a detailed exposition, see \citep{hassard1981} or \citep{raina2005}. The analysis relies on the linear, quadratic and cubic terms in the Taylor series expansion of (\ref{eq:pfmodel}), whose non-zero coefficients have been tabulated in Table \ref{tab:tayexp}. 

\begin{table}[hbtp!]
\caption{Coefficients of linear and higher order terms in the Taylor series expansion of~(\ref{eq:pfmodel}).}\label{tab:tayexp}
\renewcommand{\arraystretch}{2.25}
\centering
\begin{tabular}{ll}
\hline
\hline
$\text{Coefficients} \qquad \qquad \qquad \qquad \qquad $ & $\text{Expressions} $\\
\hline
$\xi_y=\xi_z$ & $ \dfrac{-1}{p^*} $\\  
$\xi_{yy}=\xi_{zz} $ & $ \dfrac{1}{(p^*)^2} $\\
$\xi_{xy}=\xi_{xz}$ & $ \dfrac{-1}{(p^*)^2}$\\
$\xi_{yyy}=\xi_{zzz}$ & $ \dfrac{-1}{(p^*)^3}$\\
$\xi_{xyy}=\xi_{xzz}$ & $ \dfrac{1}{(p^*)^3}$\\
\vspace{-10mm}\\
\hline
\hline
\end{tabular}
\\
\end{table}

All the notations closely follow \citep{hassard1981}. For now, we will only be concerned with the first local Hopf bifurcation. As outlined in the Appendix, the type of Hopf bifurcation and the direction of the bifurcating limit cycles can be determined from the sign of first Lyapunov coefficient ($\mu_2$) and Floquet exponent ($\beta_2$), where
\begin{eqnarray}
\mu_2 &&\hspace{-6mm}= \dfrac{-\mathbf{Re}(c_1(0))}{\alpha'(0)},\quad \beta_2= 2\mathbf{Re}(c_1(0)).\nonumber
\end{eqnarray}
We have already shown that 
\begin{equation}
\alpha'(0) =  \mathbf{Re}\left(\dfrac{d\lambda}{d\eta}\right)_{\eta=\eta_c} > 0 \quad \text{for all values of}\, \tau_1\, \text{and}\, \tau_2. 
\end{equation}
 Therefore the sign of $\mu_2$ and $\beta_2$ depends only on the sign of $\mathbf{Re}(c_1(0))$.
Using the definitions outlined in the Appendix and the values from Table \ref{tab:tayexp}, the expression for $\mathbf{Re}(c_1(0))$ for (\ref{eq:pfmodel}) has been calculated as
\begin{align}
\mathbf{Re}\big(c_1(0)\big) = &  \frac{-2\kappa\eta \sin(\vartheta) {\lvert D \rvert}^2}{{p^*}^3(1+4\sin^4(\vartheta))}\bigg((\pi/2-\vartheta)\cos(2\vartheta)\cos(\vartheta) + \cos(2\vartheta)\sin(\vartheta)  \nonumber \\
&+ \frac{\pi}{2}\big({\cos^2(2\vartheta)}+2\sin^2(\vartheta)\big)\bigg),
\end{align}
where
\begin{align*}
\vartheta &=\  \omega_0\tau_1 = \dfrac{\pi\tau_1}{\left(\tau_1+\tau_2\right)} \in (0,\pi) \quad \forall \ \tau_1, \tau_2 > 0,\\
D &= \frac{1}{1+\eta\kappa\tau_1\xi_ye^{i\omega_0\tau_1}+\eta\kappa\tau_2\xi_ze^{i\omega_0\tau_2}}.
\end{align*}
As $\kappa$, $\eta$ and $p^*>0$, we write
\begin{align*}
\mathop{\mathrm{sign}}\Big(\mathbf{Re}\big(c_1(0)\big)\Big) =& \mathop{\mathrm{sign}}\Big(\tilde{f}(\vartheta)\Big)
\end{align*}
where
\begin{align}
\tilde{f}(\vartheta) &= \frac{-2\sin(\vartheta) }{\big(1+4\sin^4(\vartheta)\big)}\bigg((\pi/2-\vartheta)\cos(2\vartheta)\cos(\vartheta)+ \cos(2\vartheta)\sin(\vartheta) \\ \nonumber &+ \frac{\pi}{2}\big({\cos^2(2\vartheta)}+2\sin^2(\vartheta)\big)\bigg).
\end{align}
\begin{figure}[hbtp!]
\center
\psfrag{x}{$\vartheta$}
\psfrag{y}[c]{$\tilde{f}(\vartheta)$}
\psfrag{0.000}{\small{\hspace{2.5mm}$0$}}
\psfrag{0.785}{\small{\hspace{1mm}$\pi/4$}}
\psfrag{1.570}{\small{\hspace{1mm}$\pi/2$}}
\psfrag{2.356}{\small{$3\pi/4$}}
\psfrag{3.141}{\small{\hspace{2.5mm}$\pi$}}
\psfrag{beta}{\small{$\beta_2$}}
\psfrag{myu2}{\small{$\mu_2$}}
\psfrag{105}{{\small{$\times 10^{-5}$}}}
\psfrag{107}{{\small{$\times 10^{-7}$}}}
\includegraphics[trim=0cm 0cm 0cm 1cm, clip=true,width=3.25in,height=2.25in]{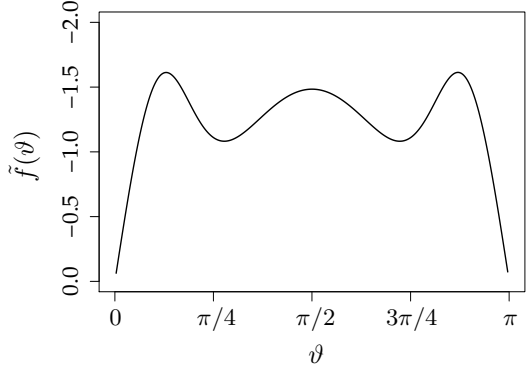}
\vspace{-2mm}
\caption{ The plot of $\tilde{f}(\vartheta)$ as $\vartheta$ varies. Note that $\tilde{f}(\vartheta)<0$, and hence $\mathbf{Re}\big(c_1(0)\big)<0$ for all values of $\vartheta \in (0,\pi)$. Therefore, the Hopf bifurcation is super-critical and the bifurcating limit cycles are asymptotically orbitally stable. }
\label{fig:pfpfrealc10}
\end{figure} 
\\
From Fig. \ref{fig:pfpfrealc10}, we can observe that $\tilde{f}(\vartheta)$ is negative for all $\vartheta \in (0,\pi)$. Therefore, $\mathbf{Re}\big(c_1(0)\big)$ is negative for all $\vartheta \in (0,\pi)$. As $\alpha'(0)>0$ for all $\vartheta \in (0,\pi)$, we get $\mu_2>0$ and $\beta_2<0$ for all values of $\tau_1$ and $\tau_2$. Therefore, the values of link capacity (C) and the round-trip times ($\tau_1,\tau_2$) do not affect the sign of $\mu_2$ and $\beta_2$, and hence the nature of Hopf bifurcation of Proportionally fair algorithm is independent of these parameters. This would appear to imply that the Proportionally fair dual algorithm would exhibit only a super-critical Hopf bifurcation, and the emerging limit cycles are asymptotically orbitally stable.
\subsection{TCP fairness}
We now consider the model of the TCP fair dual algorithm with two delays. However, in this case, given the nature of the calculus involved, derivation of general expressions (for the quantity of interest) are extremely complex and lengthy. Therefore, we resort to numerical bifurcation analysis tools. We analyze the nature of the Hopf bifurcation by plotting the bifurcation diagrams numerically (using DDE-Biftool), and then validate it through numerical simulations (using XPPAUT \citep{xppaut2002}). We consider some special cases where we choose specific values for the system parameters.  
\newcommand{\subwdth}{\textwidth}

\begin{figure}[hbtp!]
\centering
\psfrag{R}[][][2]{Amplitude of oscillations}
\psfrag{kappa}[][][2]{Bifurcation parameter, $\eta$}
\psfrag{0.95}[][][1.7]{$0.95$}
\psfrag{1.00}[][][1.7]{$1.00$}
\psfrag{1.05}[][][1.7]{$1.05$}
\psfrag{0}[][][1.7]{$0$}
\psfrag{1.01}[][][1.7]{$1.01$}
\psfrag{1.02}[][][1.7]{$1.02$}
\psfrag{1.03}[][][1.7]{$1.03$}
\psfrag{1.04}[][][1.7]{$1.04$}
\psfrag{1.05}[][][1.7]{$1.05$}
\psfrag{0}[][][1.7]{$0$}
\psfrag{1}[][][1.7]{$1$}
\psfrag{2}[][][1.7]{$2$}
\psfrag{3}[][][1.7]{$3$}
\psfrag{4}[][][1.7]{$4$}
\psfrag{5}[][][1.7]{$5$}
\psfrag{104}{\small{$\times 10^{-4}$}}
\vspace{-5mm}
\includegraphics[scale = 0.9,trim=0cm 0cm 0cm 1.7cm, clip=true,width=3.25in,height=2in]{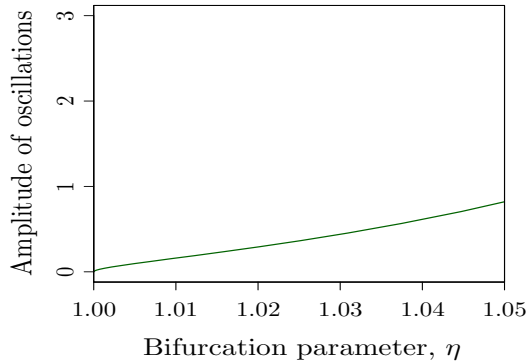}
\caption{Bifurcation diagram for the TCP fair algorithm, highlighting that the system undergoes a super-critical Hopf bifurcation for $\eta_c=1$, $\tau_1=3.2$, $\tau_2=20$ and $C= \kappa = 1$.}
\label{fig:tcptcp_bifurc_super}
\vspace{-2mm}
\end{figure}

\begin{figure}[hbtp!]
\psfrag{tcptcpsuper1}{\hspace{-3mm}{\small(a) $\eta=0.95$}}
\psfrag{tcptcpsuper2}{\hspace{-3mm}{\small(b) $\eta=1.05$}}
\psfrag{Rate}{\hspace{-5mm}\small{Price, p(t)}}
\psfrag{1000}{\hspace{-2mm}\small$1000$}
\psfrag{2000}{\hspace{-2mm}\small$2000$}
\psfrag{0.12}{\hspace{-2mm}\small$0.12$}
\psfrag{0.14}{\hspace{-2mm}\small$0.14$}
\psfrag{0.16}{\hspace{-2mm}\small$0.16$}
\psfrag{0.1}{\hspace{-2mm}\small$0.1$}
\psfrag{0.2}{\hspace{-2mm}\small$0.2$}
\psfrag{0.00}{\hspace{-0mm}\small$0$}
\psfrag{0.0}{\hspace{-1mm}\small$0$}
\psfrag{0}{\hspace{-0mm}\small$0$}
\psfrag{0.25}{\hspace{-2mm}\small $0.5$}
\psfrag{0.50}{\hspace{-2mm}\small$1.0$}
\begin{tabular}{c} 
\subfloat{\includegraphics[height=1.5in,width=0.5\subwdth]{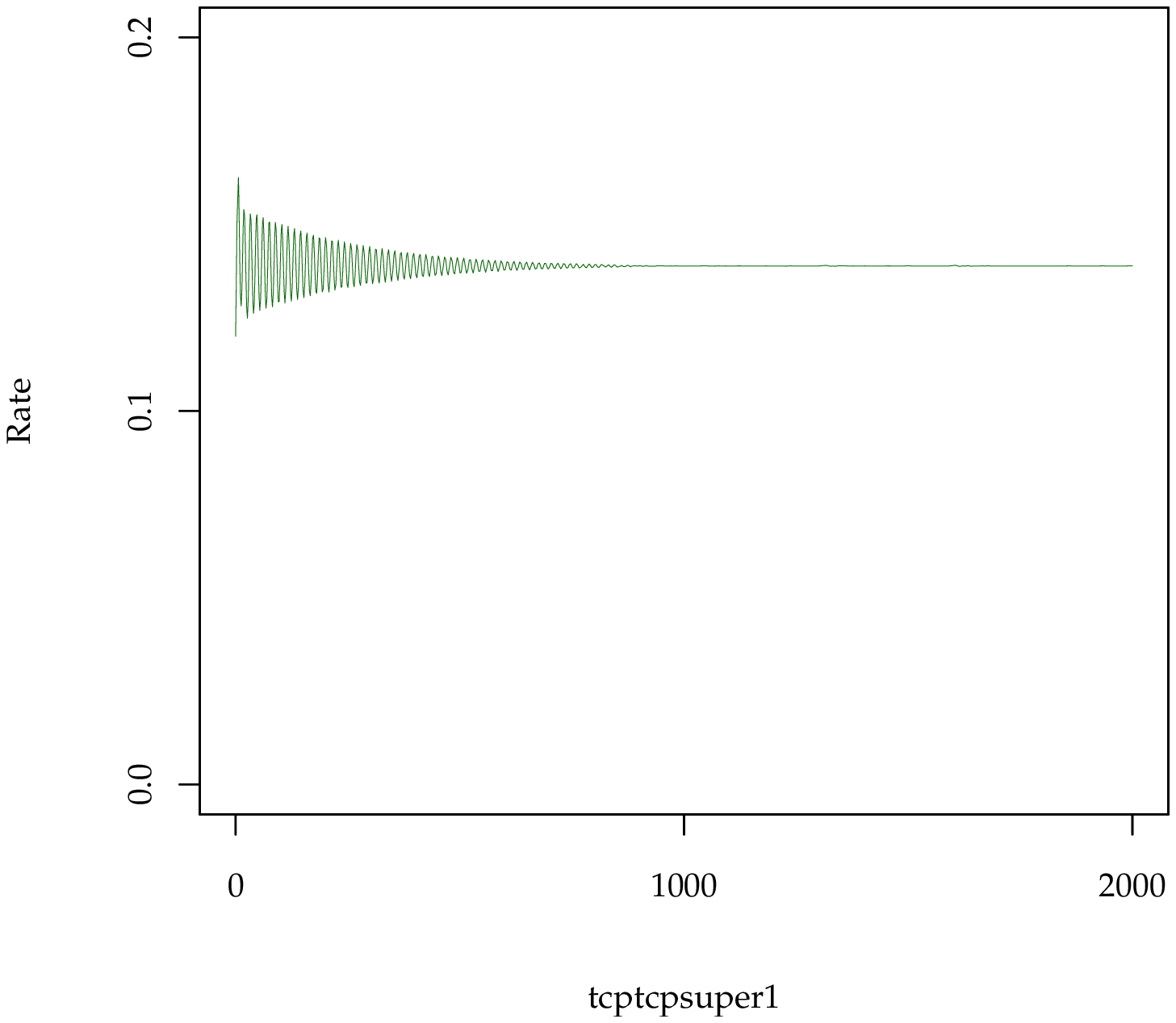}} \hspace{-10mm} 
\subfloat{\includegraphics[height=1.5in,width=0.5\subwdth]{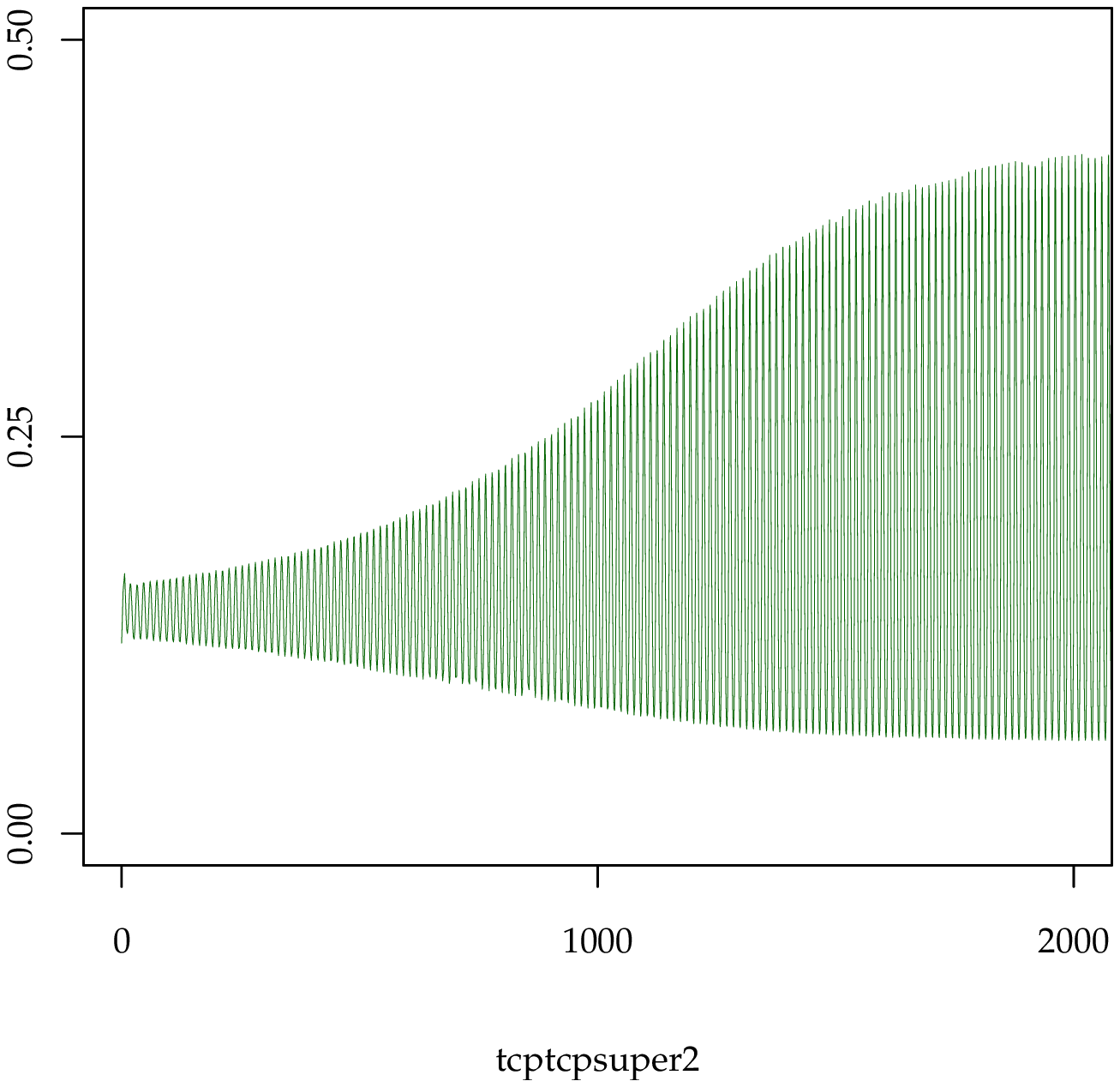}}\\
\small{Time}
\end{tabular}
\caption{Numerical simulations of the system with TCP fair allocation to illustrate the occurrence of a super-critical Hopf bifurcation as $\eta$ increases beyond the threshold. Time series are shown for the case $\tau_1=3.2$, $\tau_2=20$, $C=\kappa=1$.}
\label{fig:tcprateplot1}
\end{figure}

\textit{Numerical Example 1~(super-critical)}: Let $\tau_2=20$ and $C= \kappa = 1$, for these values the TCP fair algorithm loses stability via a Hopf bifurcation at $\tau_1=3.1$ for the critical threshold $\eta_c = 1$. We plot the bifurcation diagram (using DDE-Biftool) as shown in Fig. \ref{fig:tcptcp_bifurc_super}. We can note that as the bifurcation parameter $\eta$ increases beyond the threshold value ($\eta_c=1$), the system exhibits a super-critical Hopf bifurcation and the emerging limit cycles are stable. To validate the occurrence of super-critical Hopf, we present some numerical simulations in Fig. \ref{fig:tcprateplot1}. For $\eta=0.95$, the system converges to the stable equilibrium (see Fig. \ref{fig:tcprateplot1}(a)). Whereas, after the bifurcation i.e. for $\eta > \eta_c$, the previously stable fixed point becomes unstable and leads to the emergence of stable limit cycle (Fig. \ref{fig:tcprateplot1}(b)).

\begin{figure}[hbtp!]
\centering
\psfrag{R}[][][2]{Amplitude of oscillations}
\psfrag{kappa}[][][2]{Bifurcation parameter, $\eta$}
\psfrag{0.95}[][][1.7]{$0.95$}
\psfrag{0.98}[][][1.7]{$0.98$}
\psfrag{0.99}[][][1.7]{$0.99$}
\psfrag{1.00}[][][1.7]{$1.00$}
\psfrag{1.01}[][][1.7]{$1.01$}
\psfrag{1.02}[][][1.7]{$1.02$}
\psfrag{1.03}[][][1.7]{$1.03$}
\psfrag{1.04}[][][1.7]{$1.04$}
\psfrag{1.05}[][][1.7]{$1.05$}
\psfrag{0}[][][1.7]{$0$}
\psfrag{0.0}[][][1.7]{$0.0$}
\psfrag{2.5}[][][1.7]{$2.5$}
\psfrag{5.0}[][][1.7]{$5.0$}
\psfrag{1}[][][1.7]{$1$}
\psfrag{2}[][][1.7]{$2$}
\psfrag{3}[][][1.7]{$3$}
\psfrag{4}[][][1.7]{$4$}
\psfrag{5}[][][1.7]{$5$}
\psfrag{6}[][][1.7]{$6$}
\psfrag{9}[][][1.7]{$9$}
\psfrag{10}[][][1.7]{$10$}
\psfrag{60}[][][1.7]{$60$}
\psfrag{104}{\small{$\times 10^{-4}$}}
\vspace{-4mm}
\includegraphics[scale = 0.9,trim=0cm 0cm 0cm 1.7cm, clip=true,width=3.25in,height=2in]{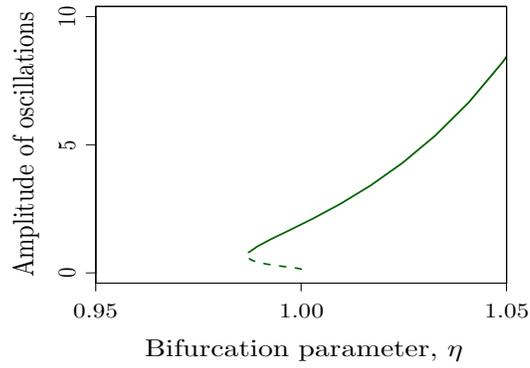}
\caption{Bifurcation diagram for the TCP fair algorithm, highlighting the existence of a sub-critical Hopf for $\eta_c=1$, $\tau_1=3.7$, $\tau_2=10$, $C=1$.}
\label{fig:tcptcp_bifurc_sub}
\end{figure}

\begin{figure}[hbtp!]
\centering
\psfrag{tcptcpsub1}{\hspace{-3mm}{\small(a) $\eta=0.95$}}
\psfrag{tcptcpsub2}{\hspace{-3mm}{\small(b) $\eta=1.05$}}
\psfrag{Rate}{\hspace{-5mm}\small{Price, p(t)}}
\psfrag{1000}{\hspace{-2mm}\small$1000$}
\psfrag{2000}{\hspace{-2mm}\small$2000$}
\psfrag{0.11}{\hspace{-2mm}\small$0.11$}
\psfrag{0.14}{\hspace{-2mm}\small$0.14$}
\psfrag{0.1}{\hspace{-2mm}\small$0.1$}
\psfrag{0.2}{\hspace{-2mm}\small$0.2$}
\psfrag{0.17}{\hspace{-2mm}\small$0.17$}
\psfrag{0.00}{\hspace{-0mm}\small$0$}
\psfrag{0.0}{\hspace{-1mm}\small$0$}
\psfrag{0}{\hspace{-0mm}\small$0$}
\psfrag{2.5}{\hspace{-0mm}\small$5$}
\psfrag{5.0}{\hspace{-2mm}\small$10$}
\begin{tabular}{c}
\subfloat{\includegraphics[height=1.5in,width=0.5\subwdth]{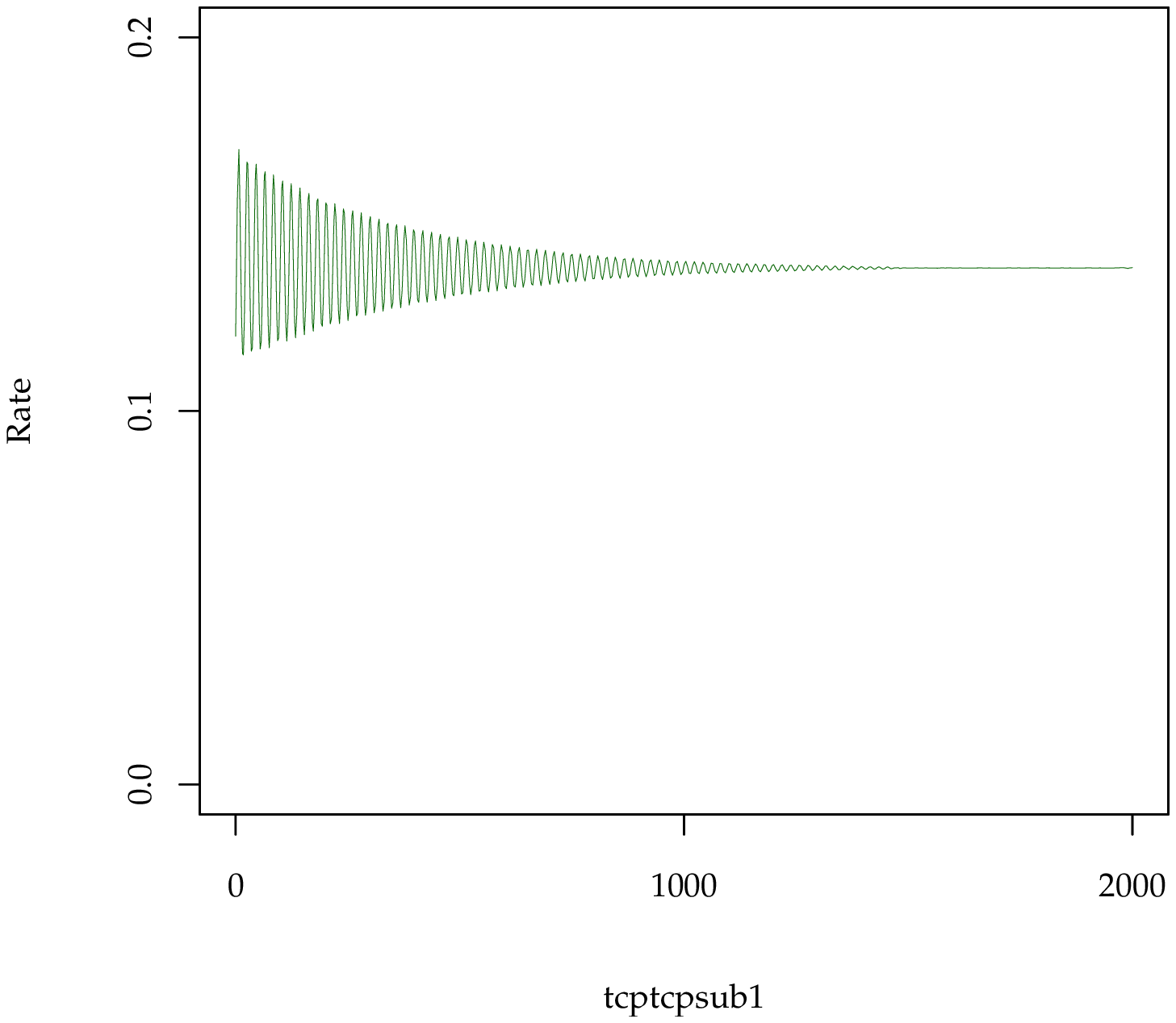}} \hspace{-10mm} 
\subfloat{\includegraphics[height=1.5in,width=0.5\subwdth]{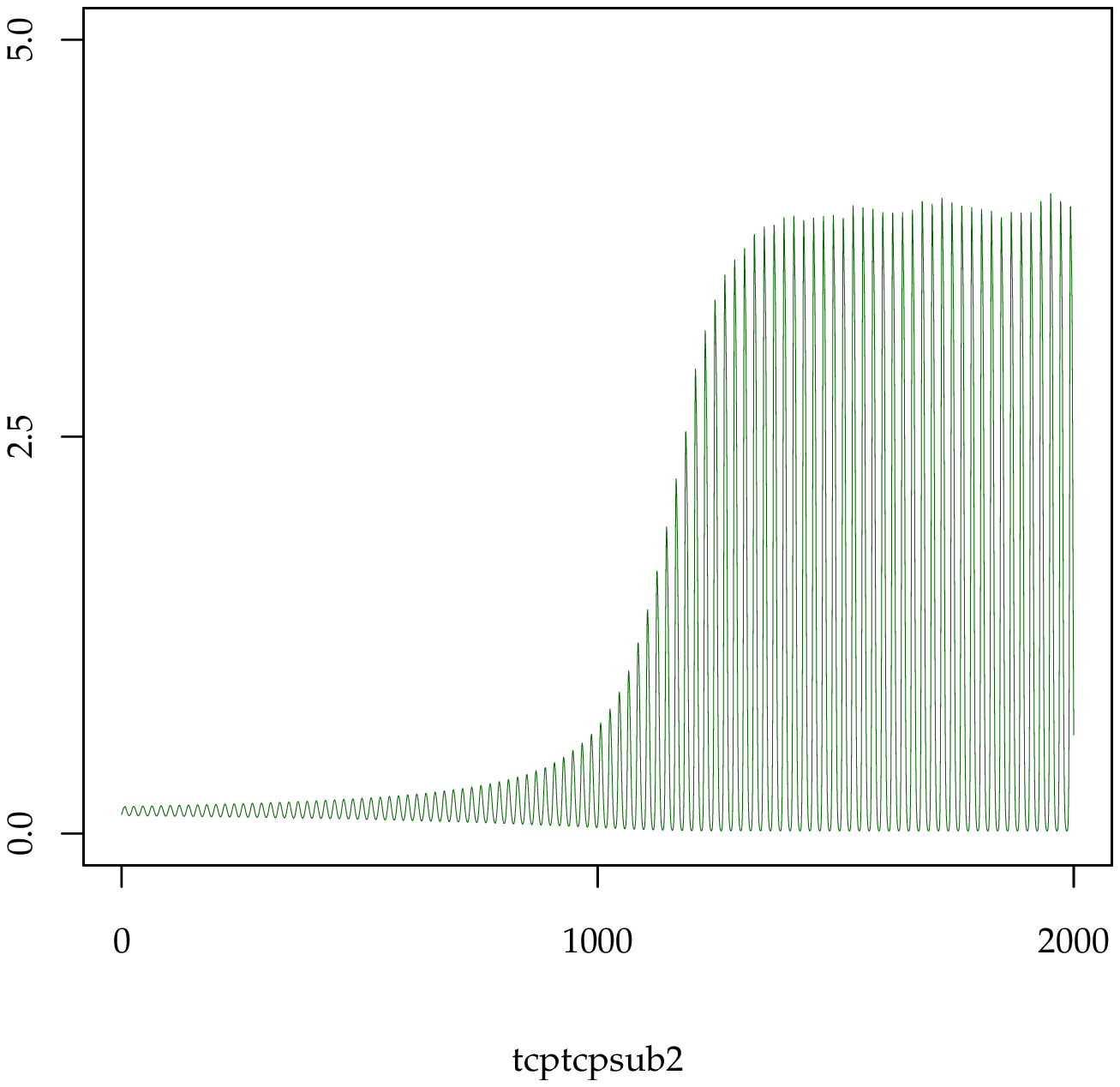}}\\
\small{Time}
\end{tabular}
\caption{Numerical simulations of the system with TCP fair allocation to validate the occurrence of sub-critical Hopf bifurcation as $\eta$ increases beyond $\eta_c$. The parameter values chosen are $\tau_1=3.7$, $\tau_2=10$ and $\kappa = C = 1$.}
\label{fig:tcprateplot2}
\end{figure}

\textit{\ \ Numerical Example 2~(sub-critical)}: For the parameter values $\tau_1=3.7$, $\tau_2=10$ and $C=1$, the system loses local stability at $\eta$ = 1. 
As shown in Fig.~\ref{fig:tcptcp_bifurc_sub}, the system undergoes a sub-critical Hopf bifurcation which results in the emergence of large amplitude limit cycles. The solid and dashed lines denote the amplitude of stable and unstable limit cycles respectively. The numerical simulation shown in Fig.~\ref{fig:tcprateplot2}(a) illustrates that the system is locally stable for $\eta<1$. But, after the occurrence of bifurcation, the system converges to a large amplitude limit cycle (Fig.~\ref{fig:tcprateplot2}(b)). Thus, in the case of two delays, the TCP fair algorithm can undergo a sub-critical Hopf bifurcation, for some parameter values.


\subsection{Delay dual}
We now consider the Delay dual algorithm with two delays. Again, given the nature of analysis, we consider some specific parameter values, and proceed with the DDE-Biftool.

\textit{ Numerical Example 1 (super-critical)}: For $\tau_1=20$, $\tau_2=10$, $\alpha_s=1.84$ and $C=1$, the system undergoes a Hopf bifurcation at $\eta = \eta_c = 1$.
\begin{figure}[h!]
\centering
\psfrag{R}[][][2]{Amplitude of oscillations}
\psfrag{kappa}[][][2]{Bifurcation parameter, $\eta$}
\psfrag{10}[][][1.7]{$10$}
\psfrag{1.00}[][][1.7]{$1.00$}
\psfrag{1.05}[][][1.7]{$1.05$}
\psfrag{0}[][][1.7]{$0$}
\psfrag{1.01}[][][1.7]{$1.01$}
\psfrag{1.02}[][][1.7]{$1.02$}
\psfrag{1.03}[][][1.7]{$1.03$}
\psfrag{1.04}[][][1.7]{$1.04$}
\psfrag{1.05}[][][1.7]{$1.05$}
\psfrag{20}[][][1.7]{$20$}
\psfrag{30}[][][1.7]{$30$}
\psfrag{40}[][][1.7]{$40$}
\psfrag{104}{\small{$\times 10^{-4}$}}
\psfrag{0.0}[][][1.7]{$0.0$}
\psfrag{2.5}[][][1.7]{$2.5$}
\psfrag{5.0}[][][1.7]{$5.0$}
\vspace{-2mm}
\includegraphics[scale = 0.9,trim=0cm 0cm 0cm 1.7cm, clip=true,width=3.25in,height=2in]{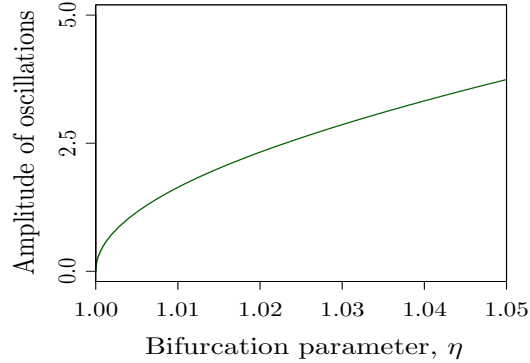}
\caption{Bifurcation diagram for the Delay dual algorithm, showing the existence of super-critical Hopf for $\tau_1=5$, $\tau_2=30$, $\alpha_s=4.0$ and $C=1$.}
\label{fig:dd_bifurc_super}
\end{figure}
\begin{figure}[hbtp!]
\psfrag{ddsuper1}{\hspace{-1mm}{\small(a) $\eta=0.95$}}
\psfrag{ddsuper2}{\hspace{-1mm}{\small(b) $\eta=1.05$}}
\psfrag{Rate}{\hspace{-5mm}\small{Price, p(t)}}
\psfrag{750}{\hspace{-2mm}\small$750$}
\psfrag{1500}{\hspace{-2mm}\small$1500$}
\psfrag{1.6}{\hspace{-2mm}\small$1.6$}
\psfrag{1.8}{\hspace{-2mm}\small$1.8$}
\psfrag{1}{\hspace{-2mm}\small$1.25$}
\psfrag{2.0}{\hspace{-2mm}\small$2.0$}
\psfrag{0.00}{\hspace{-0mm}\small$0$}
\psfrag{0.0}{\hspace{-1mm}\small$0$}
\psfrag{0}{\hspace{-1mm}\small$0$}
\psfrag{2}{\hspace{-1mm}\small$2.5$}
\psfrag{4}{\hspace{-1mm}\small$5.0$}
\psfrag{2000}{\hspace{-2mm}\small$2000$}
\psfrag{4000}{\hspace{-2mm}\small$4000$}
\begin{tabular}{c} 
\subfloat{\includegraphics[height=1.5in,width=0.5\subwdth]{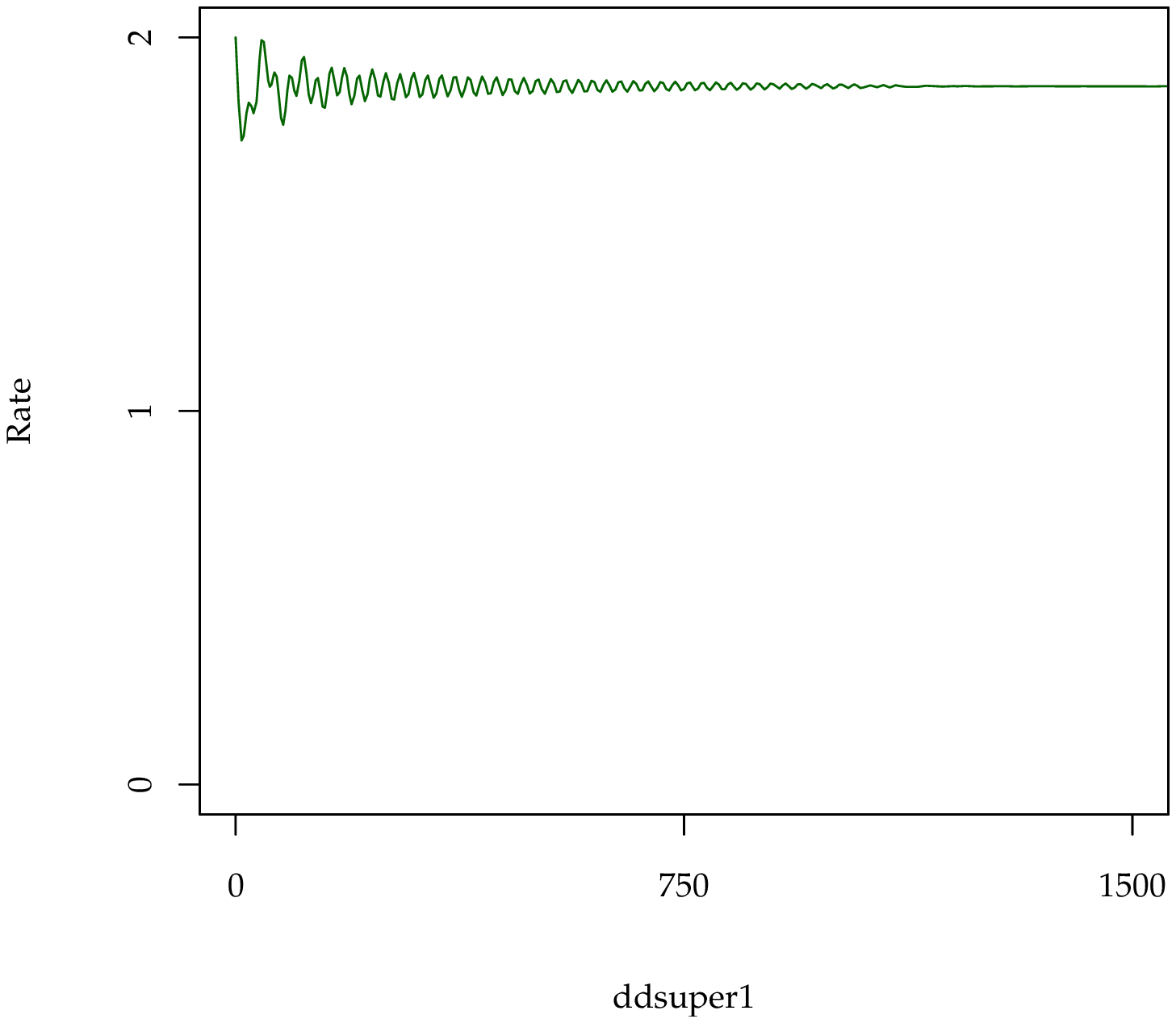}} \hspace{-10mm} 
\subfloat{\includegraphics[height=1.5in,width=0.5\subwdth]{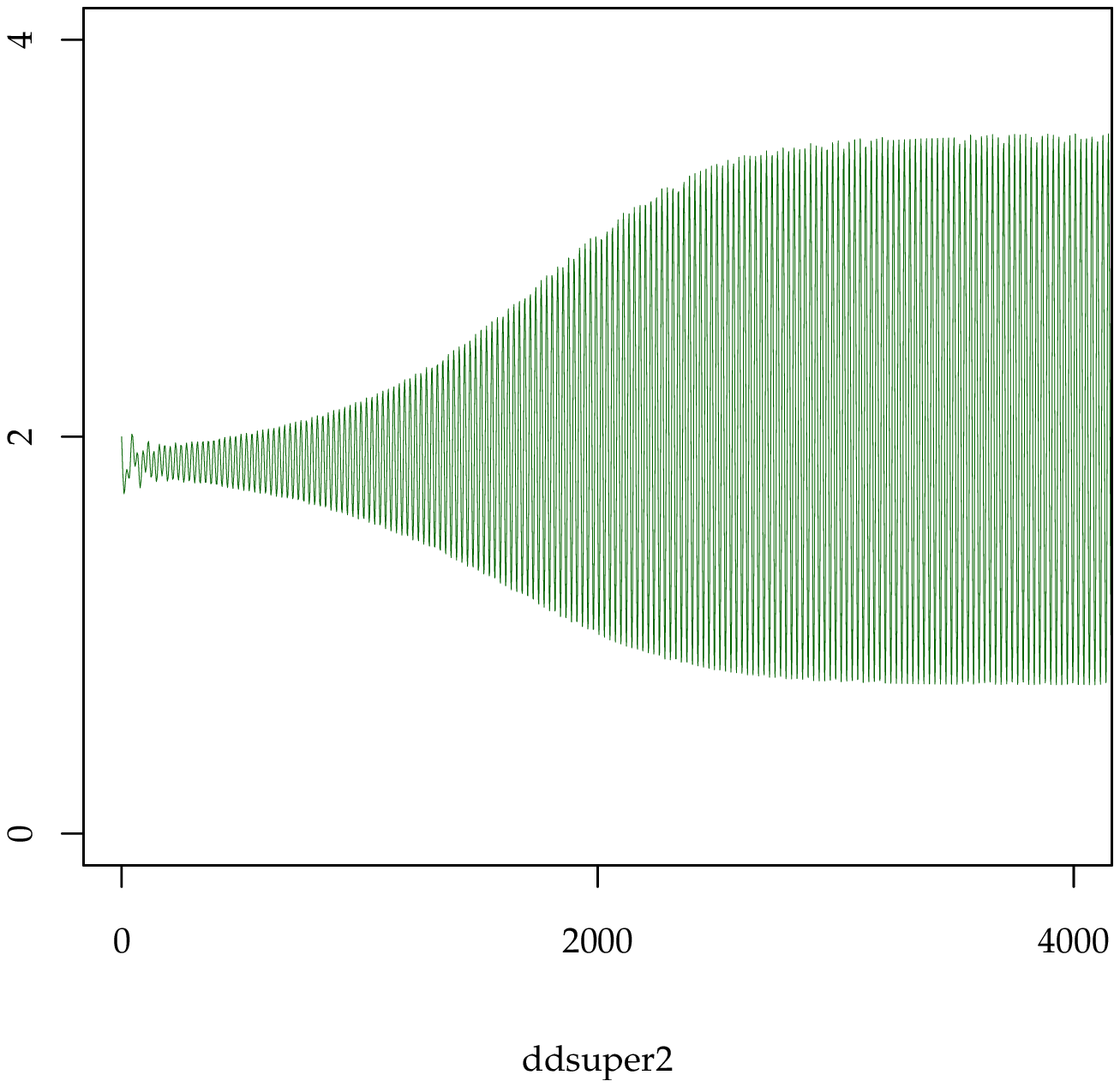}}\\
\small{Time}
\end{tabular}
\caption{Numerical simulations of the system with Delay dual algorithm to illustrate the existence of a super-critical Hopf bifurcation for $\tau_1=5$, $\tau_2=30$, $\alpha_s = 4$ and $C=\kappa=1$.}
\label{fig:ddrateplot1}
\end{figure}
From Fig.~\ref{fig:dd_bifurc_super} we can see that, as $\eta$ is varied beyond $\eta_c$, the system exhibits a super-critical Hopf and gives rise to small amplitude stable limit cycles. The numerical simulations shown in Fig.~\ref{fig:ddrateplot1}(a) illustrate that the system is locally stable for $\eta<1$. But, after the occurrence of bifurcation, the system converges to a stable limit cycle (Fig.~\ref{fig:ddrateplot1}(b)).

\begin{figure}[h!]
\centering
\psfrag{R}[][][2]{Amplitude of oscillations}
\psfrag{kappa}[][][2]{Bifurcation parameter, $\eta$}
\psfrag{0.95}[][][1.6]{$0.95$}
\psfrag{1.01}[][][1.6]{$1.01$}
\psfrag{1.05}[][][1.6]{$1.05$}
\psfrag{0.96}[][][1.6]{$0.96$}
\psfrag{0.97}[][][1.6]{$0.97$}
\psfrag{0.98}[][][1.6]{$0.98$}
\psfrag{0.99}[][][1.6]{$0.99$}
\psfrag{1.00}[][][1.6]{$1.00$}
\psfrag{1.02}[][][1.6]{$1.02$}
\psfrag{1.03}[][][1.6]{$1.03$}
\psfrag{1.04}[][][1.6]{$1.04$}
\psfrag{0}[][][1.6]{$0$}
\psfrag{100}[][][1.6]{$100$}
\psfrag{20}[][][1.6]{$20$}
\psfrag{30}[][][1.6]{$30$}
\psfrag{40}[][][1.6]{$40$}
\psfrag{50}[][][1.6]{$50$}
\psfrag{150}[][][1.6]{$150$}
\psfrag{45}[][][1.6]{$45$}
\psfrag{60}[][][1.6]{$60$}
\psfrag{104}{\small{$\times 10^{-4}$}}
\psfrag{0.0}[][][1.7]{$0.0$}
\psfrag{2.5}[][][1.7]{$2.5$}
\psfrag{5.0}[][][1.7]{$5.0$}
\vspace{-2mm}
\includegraphics[scale = 0.9,trim=0cm 0cm 0cm 1.7cm, clip=true,width=3.25in,height=2in]{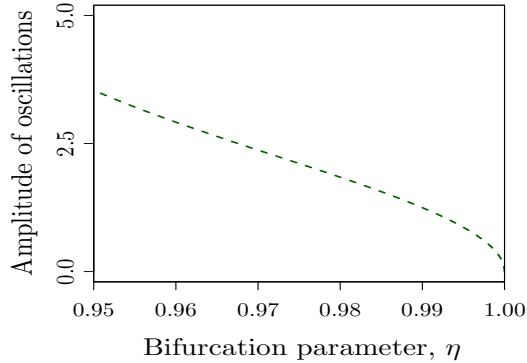}
\caption{Bifurcation diagram for the Delay dual algorithm, showing that the system undergoes a sub-critical Hopf for $\tau_1=5$, $\tau_2=40$, $\kappa=1.2$ $\alpha_s=4.1$ and $C=1$.}
\label{fig:dd_bifurc_sub}
\end{figure}

\begin{figure}[hbtp!]
\psfrag{ddsub1}{\hspace{-1mm}{\small(a) $\eta=0.95$}}
\psfrag{ddsub2}{\hspace{-1mm}{\small(b) $\eta=1.05$}}
\psfrag{Rate}{\hspace{-5mm}\small{Price, p(t)}}
\psfrag{100}{\hspace{-2mm}\small$100$}
\psfrag{1500}{\hspace{-2mm}\small$1500$}
\psfrag{1.6}{\hspace{-2mm}\small$1.6$}
\psfrag{1.8}{\hspace{-2mm}\small$1.8$}
\psfrag{2.0}{\hspace{-2mm}\small$2.0$}
\psfrag{0.00}{\hspace{-0mm}\small$0$}
\psfrag{0.0}{\hspace{-1mm}\small$0$}
\psfrag{0}{\hspace{-1mm}\small$0$}
\psfrag{2.4}{\hspace{-1mm}\small$2.4$}
\psfrag{50}{\hspace{-1mm}\small$50$}
\psfrag{3000}{\hspace{-2mm}\small$3000$}
\psfrag{4000}{\hspace{-2mm}\small$4000$}
\psfrag{1.25}{\hspace{-2mm}\small$1.25$}
\psfrag{2.50}{\hspace{-2mm}\small$2.50$}
\psfrag{1.5}{\hspace{-2mm}\small$1.5$}
\psfrag{3.0}{\hspace{-2mm}\small$3.0$}

\begin{tabular}{c}
\subfloat{\includegraphics[height=1.5in,width=0.5\subwdth]{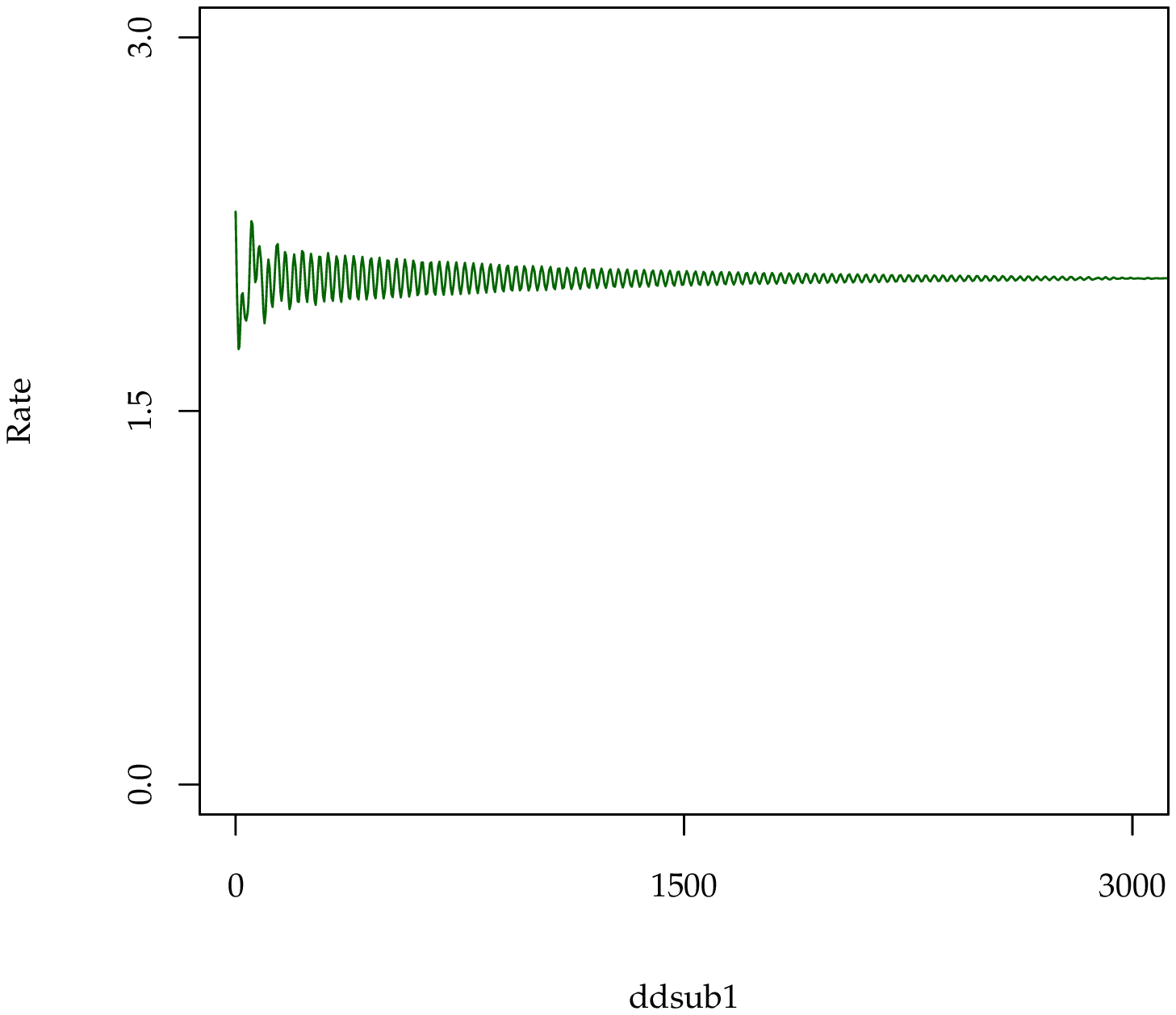}} \hspace{-10mm} 
\subfloat{\includegraphics[height=1.5in,width=0.5\subwdth]{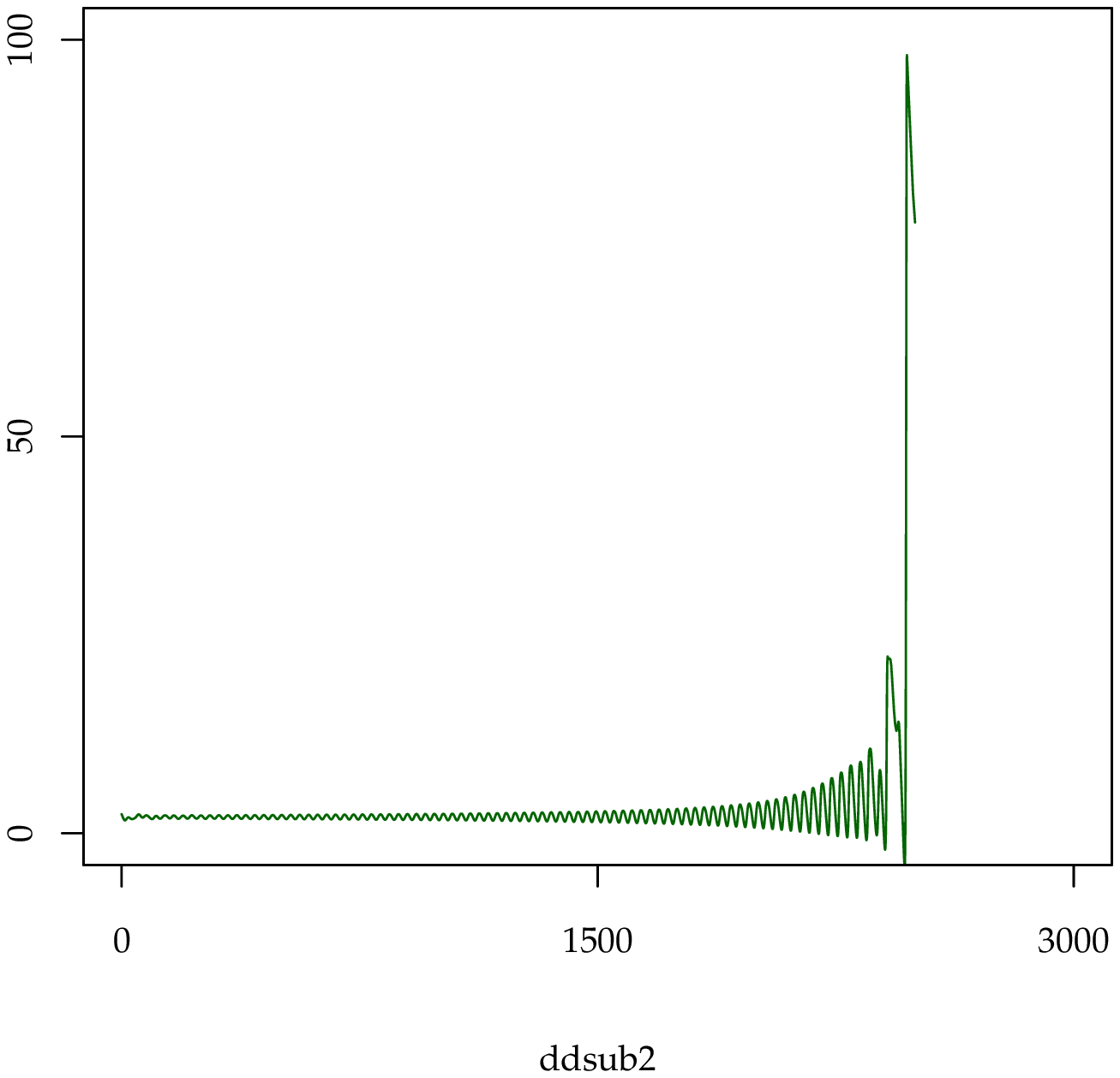}}\\
\small{Time}
\end{tabular}
\caption{Numerical simulations for the Delay dual algorithm highlighting that the system exhibit a sub-critical Hopf for the parameter values $\alpha_s=4.1$, $\kappa=1.2$, $C = 1$, $\tau_1=5$ and $\tau_2=40$.}
\label{fig:ddrateplot2}
\end{figure}

\textit{ Numerical Example 2 (sub-critical)}: Consider $\tau_1=5$, $\tau_2=40$, $\alpha_s=4.1$ and $C=1$. As shown in Fig.~\ref{fig:dd_bifurc_sub}, the system undergoes a sub-critical Hopf bifurcation at $\eta=1$. From Fig. \ref{fig:ddrateplot2}(a), we can verify that the system converges to the equilibrium for $\eta < \eta_c$. Whereas, after the bifurcation i.e. for $\eta > \eta_c$, the previously stable fixed point becomes unstable and also the solution would eventually jump to infinity (Fig. \ref{fig:ddrateplot2}(b)). 

The results of Hopf bifurcation analysis would seem to provide evidence 
that the Proportionally fair algorithm would undergo only super-critical Hopf bifurcation, and the bifurcating limit cycles are orbitally asymptotically stable. Whereas, the TCP fair and Delay dual algorithms can undergo a sub-critical Hopf bifurcation. In general, a sub-critical Hopf bifurcation would give rise either to the onset of large amplitude limit cycles or to unstable limit cycles \citep{strogatz2018}. In the design of congestion control algorithms in which there is a possibility of violation of the conditions for local stability, an important design objective is not only to make sure that the system is stable, but also to ensure any loss of system stability produces stable limit cycles with small amplitude. Therefore, the results of our analyses tend to favor the Proportionally fair allocation in dual algorithms.
\\ \\
\textit{Impact of delay heterogeneity on the nature of the Hopf bifurcation}: As highlighted in Table \ref{tab:12delayhopf}, in the case of single delay, all the three algorithms always exhibit a super-critical Hopf bifurcation. In the case of two delay variants, analytical evidence seems to imply that the Proportionally fair algorithm would produce only a super-critical Hopf bifurcation. Whereas, the TCP fair and Delay dual algorithms can undergo a sub-critical Hopf bifurcation, for some parameter values. Therefore, both the delays do appear to play an important role in determining the type of the Hopf bifurcation. We believe that this work is an important step towards understanding the impact of delay heterogeneity on the dynamics of dual congestion control algorithms.

\begin{table}[hbtp!]
\caption{Comparison of the Hopf bifurcation results between the single delay and two delay variants of dual algorithms. Observe that, in the case of TCP fair and Delay dual algorithms, the delay heterogeneity affects the nature of Hopf bifurcation.}\label{tab:12delayhopf}
\resizebox{\columnwidth}{!}{
\begin{tabular}{ll}
\hline
\hline
$\textbf{Algorithm}    $  & $ \hspace{-20mm}\textbf{Nature of Hopf bifurcation} $ \vspace{-0mm}\\ 
$\qquad \qquad  \qquad \qquad \qquad \hspace{5mm} \textbf{Single delay}$ $ \hspace{-7mm}$  & $\hspace{5mm}\textbf{Two delays}$\\
\hline
\vspace{1mm}\\
\text{Proportional fairness} $   $ $\hspace{2mm}\text{always super-critical }$ & $ \text{evidence suggests super-critical } $  \\  \\ \vspace{-0mm}
\text{TCP fairness} $   \quad \,   $ $ \hspace{2mm}\qquad \text{always super-critical}$ & $\text{evidence suggests sub-critical }$\\ \\
\text{Delay dual} $  \,  \qquad$ $ \qquad \hspace{2mm}  \text{always super-critical }$ & $ \text{evidence suggests sub-critical}$\\   \\
\hline 
\hline
\end{tabular}
}
\\
\end{table}

\section{Contributions}
Fairness and stability are two key considerations in the design of congestion control algorithms. Some of the important notions of fairness, in the class of dual algorithms, are: i) Proportional fairness, ii) TCP fairness, and iii) Delay-based fairness. In this paper, we considered dual congestion control models coupled with these three different notions of fairness, and analyzed the stability and the consequences associated with the loss of stability. From a stability perspective, it is possible to derive sufficient conditions that will ensure local stability for all the three notions of fairness. While stability is important, the models are non-linear and so it seems useful to explore some non-linear properties of these models. In particular, we proceeded to analyze the dynamics of these algorithms as conditions for stability are just violated. We found that the loss of stability happens via a Hopf bifurcation, and we outlined an analytical framework to establish the type of the Hopf bifurcation, and the stability of the emerging limit cycles. 

We showed numerically that the TCP fair and the Delay dual algorithms can undergo a sub-critical Hopf bifurcation, which is undesirable for engineering applications. In contrast, in the case of the Proportionally fair algorithm, we provided strong evidence to suggest that the occurrence of a super-critical Hopf bifurcation is what we should expect as the stability condition gets violated. Therefore, the results of our analysis tend to favor the Proportionally fair allocation in the class of dual algorithms. Another interesting comparison can be made between the results for the single-delay and two-delay variants of these algorithms. In the case of single delay, the type of Hopf bifurcation for all the three algorithms is always super-critical. Whereas for the two delay variants, evidences suggest that the Proportionally fair algorithm exhibit only a super-critical Hopf, and the TCP fair and Delay dual algorithms can undergo a sub-critical Hopf for some parameter values. We also analyzed a scalar non-linear equation with two discrete delays (see Example (A.1) in the Appendix), and showed that it \textit{may} be possible to switch the type of the bifurcation by simply changing the delays. 	

In the context of dual algorithms, this is the first study that highlights the existence of a sub-critical Hopf bifurcation, which we believe has been previously overlooked. One may also gain some additional insights if the analysis can be extended to multi-link topologies with multiple delays.
\label{apndx}
\section*{Appendix}
We analyze an autonomous non-linear equation with two discrete delays. We first derive some sufficient conditions for local stability. Following the style of analysis in \citep{hassard1981} (also see \citep{raina2005} for a detailed exposition), we then outline the necessary calculations to determine the type of Hopf bifurcation and the asymptotic orbital stability of the
bifurcation solutions as local instability just sets in. We opt to use a non-dimensional parameter, $\eta$, as the bifurcation parameter. For now,
we will only be concerned with the first Hopf bifurcation. The
framework employed to address the stability of the limit cycles
is the Poincar{\'e} normal forms and the Center manifold theorem.
Consider the following non-linear delay differential equation
\begin{equation}
\label{eq:gen_non_eq}
\dfrac{d}{dt}x(t) = \eta f\big(x(t),x(t-\tau_1),x(t-\tau_2)\big),
\end{equation}
where $f$ has a unique equilibrium denoted by $(x^*,y^*,z^*)$ and $\tau_1,\ \tau_2,\  \eta > 0$. Define $u(t) = x(t)-x^*,$ and take a Taylor series expansion for  (\ref{eq:gen_non_eq}) including the linear, quadratic and cubic terms to obtain
\begin{equation}
\begin{aligned}
\dfrac{d}{dt}u(t)=\  &\eta \big(\xi_yu(t-\tau_1)+\xi_zu(t-\tau_2)+\xi_{xy}u(t)u(t-\tau_1)+\xi_{xz}u(t)u(t-\tau_2)\\
&+\xi_{yy}u^2(t-\tau_1)+\xi_{yz}u(t-\tau_1)u(t-\tau_2)+\xi_{zz}u^2(t-\tau_2)\\
&+\xi_{xyy}u(t)u^2(t-\tau_1)+\xi_{xzz}u(t)u^2(t-\tau_2)+\xi_{yyy}u^3(t-\tau_1)\\
&+\xi_{xyz}u(t)u(t-\tau_1)u(t-\tau_2)+\xi_{yyz}u^2(t-\tau_1)u(t-\tau_2)\\
&+\xi_{zzz}u^3(t-\tau_2)+\xi_{yzz}u(t-\tau_1)u^2(t-\tau_2)+ \mathcal{O}(u^4)\big)
\end{aligned}
\label{eq:linear_noneq}
\end{equation}
where, letting $f^*$ denote evaluation of $f$ at $(x^*,y^*,z^*)$
\begin{alignat*}
\xi\xi_i&=f^*_i,&\xi_{ii}&=\dfrac{1}{2}f^*_{ii},&\xi_{iii}&=\dfrac{1}{6}f^*_{iii} &\forall \  i &\in \{x,y,z\} & \ \\
\xi_{xy}&=f^*_{xy},& \xi_{xz}&=f^*_{xz},& \xi_{yz}&=f^*_{yz},& \xi_{xxy}&=\dfrac{1}{2}f^*_{xxy},& \xi_{xyz}&=f^*_{xyz},\\
\xi_{xxz}&=\dfrac{1}{2}f^*_{xxz},\ \  & \xi_{xyy}&=\dfrac{1}{2}f^*_{xyy},\ \  &\xi_{xzz}&=\dfrac{1}{2}f^*_{xzz}, \ \  & \xi_{yyz}&=\dfrac{1}{2}f^*_{yyz},\ \  &\xi_{yzz}&=\dfrac{1}{2}f^*_{yzz}.
\end{alignat*}
We state here some results for convenience of reference.\\ \\
\textbf{Result I} Consider a linear autonomous delay equation whose corresponding characteristic equation is given by 
\begin{equation}
 \lambda + \eta a_1 e^{-\lambda \tau_1} + \eta a_2 e^{-\lambda \tau_2} = 0,
 \label{eq:apndxce1}
\end{equation}
where $\eta$, $a_1$, $a_2$, $\tau_1$, $\tau_2 > 0$, then a \textit{sufficient} condition for the trivial solution of the corresponding system to be stable is 
\begin{equation}
 \eta a_1 \tau_1 + \eta a_2 \tau_2 < 1.
\end{equation}
\textit{Calculations}. We now use Nyquist stability criterion to obtain the above stability condition. To obtain a loop transfer function, we rewrite the characteristic equation \eqref{eq:apndxce1} as 
\begin{equation}
 \lambda \Bigg(1+ \frac{\eta a_1 e^{-\lambda\tau_1}}{\lambda}+ \frac{\eta a_2 e^{-\lambda\tau_2}}{\lambda}\Bigg)=0.
 \end{equation}
Now, the loop transfer function is given 
\begin{equation}
L(\lambda)= \frac{\eta a_1 e^{-\lambda\tau_1}}{\lambda}+ \frac{\eta a_2 e^{-\lambda\tau_2}}{\lambda} 
\label{eq:apndxlooptf}
\end{equation}
The next step is to obtain the crossover frequency $\omega_c$ at which the $\phase{L(j\omega_c)} =\pi$. At this frequency, the magnitude of the loop transfer function should be less than 1, i.e., $\left|L(j\omega)\right| < 1$. Now, substituting $\lambda=j\omega$ in \eqref{eq:apndxlooptf} yields
\begin{align*}
 L(j\omega)&= \frac{\eta a_1 e^{-j\omega\tau_1}}{j\omega}+ \frac{\eta a_2 e^{-j\omega\tau_2}}{j\omega}\\ \nonumber
 &=-\frac{\eta}{\omega}\bigg(a_1 \sin(\omega\tau_1) + a_2 \sin(\omega\tau_2) + j\big(a_1 \cos(\omega\tau_1) + a_2 \cos(\omega\tau_2)\big)\bigg).
\end{align*}
Equating $\phase{L(j\omega_c)}$ to $\pi$, we get
\begin{equation}
 a_1 \cos(\omega_c\tau_1) + a_2 \cos(\omega_c\tau_2) = 0.
 \label{eq:apndx_angle_condn}
\end{equation}
Similarly, the magnitude condition $\left|L(j\omega)\right| < 1$ can be written as 
\begin{equation}
{
 \big(a_1 \sin(\omega\tau_1) + a_2 \sin(\omega\tau_2)\big)^2 + \big(a_1 \cos(\omega\tau_1) + a_2 \cos(\omega\tau_2)\big)^2 < \frac{\omega^2}{\eta^2}.
\label{eq:apndx_mag_condn1}
}
\end{equation}
Substituting \eqref{eq:apndx_angle_condn} in \eqref{eq:apndx_mag_condn1}, we obtain
\begin{equation}
 \frac{\eta a_1\sin(\omega_c\tau_1)}{\omega_c} + \frac{\eta a_2\sin(\omega_c\tau_2)}{\omega_c} < 1.
 \label{eq:apndx_mag_condn2}
\end{equation}
We can rewrite \eqref{eq:apndx_mag_condn2} as
\begin{equation}
 \eta a_1 \tau_1\frac{\sin(\omega_c\tau_1)}{\omega_c\tau_1} + \eta a_2 \tau_2\frac{\sin(\omega_c\tau_2)}{\omega_c\tau_1} < 1.
 \label{eq:apndx_mag_condn3}
\end{equation}
We know that $\left({\sin(\theta)}/{\theta}\right)\leq1$ for all values of $\theta$. Therefore, if we ensure that $\eta a_{1}\tau_1+ \eta a_{2}\tau_2 < 1$  then the condition \eqref{eq:apndx_mag_condn3} will be satisfied, and hence the system is locally asymptotically stable.\\

Now, consider the case $a_1 = a_2 = a > 0$.\\ \\
\textbf{Result II} Consider a linear autonomous delay equation whose corresponding characteristic equation is given by 
\begin{equation}
 \lambda + \eta a  e^{-\lambda \tau_1} + \eta a e^{-\lambda \tau_2} = 0,
 \label{eq:apndxce2}
\end{equation}
where $\eta$, $a$, $\tau_1$, $\tau_2 > 0$, then a \textit{sufficient} condition for stability is 
\begin{equation}
 \eta a \tau_1 + \eta a \tau_2 < 1.
\end{equation}
The above condition can be obtained by an analysis similar to that of Result I.\\

With $a_1=a$ and $a_2=0$, we can obtain the stability results for single delay.\\ \\
\textbf{Result III} Consider a linear autonomous delay equation whose corresponding characteristic equation is given by 
\begin{equation}
 \lambda + \eta a e^{-\lambda \tau} = 0,
 \label{eq:apndxce3}
\end{equation}
where $\eta$, $a$, $\tau > 0$, then a \textit{sufficient} condition for the corresponding system to be stable is 
\begin{equation}
 \eta a \tau < 1.
\end{equation}

The calculations that follow will enable us to address questions about the nature of the bifurcating solutions, as the system transits
from stability to instability via a Hopf bifurcation. For this we have to take higher order terms, i.e., the quadratic and cubic of (\ref{eq:linear_noneq}) into consideration. Following the style of analysis in \citep{raina2005}, we now do the calculations in the case of two delays.

Consider the following delay differential equation
\begin{equation}
\label{eq:auto_noneq}
\dfrac{d}{dt}u(t) = \mathcal{L}_\mu u_t + \mathcal{F}(u_t,\mu),
\end{equation}
where $t>0,\ \mu \in \mathbb{R}$ and $\tau = \max(\tau_1,\tau_2)$.
\begin{equation*}
  u_t(\theta) = u(t+\theta), \quad u:[-\tau,0]\rightarrow\mathbb{R}\ \quad \text{and} \quad \theta\in[-\tau,0].
\end{equation*}
$\mathcal{L}_\mu:C[-\tau,0]\rightarrow\mathbb{R}$ and $\mathcal{F}:C[-\tau,0]\rightarrow\mathbb{R}$ are one parameter family of linear and non-linear operators respectively. We consider that $\mathcal{F}(u_t,\mu)$ is analytic and that $\mathcal{F}$ and $\mathcal{L}_\mu$ depend analytically on the bifurcation parameter. We now rewrite (\ref{eq:auto_noneq}) into the following form
\begin{equation}
\label{eq:auto_noneq_matrix}
 \dfrac{d}{dt}u_t  = \mathcal{A}(\mu)u_t+\mathcal{R}u_t,
\end{equation}
which has $u_t$ rather than both $u$ and $u_t$. We now employ the Riesz representation theorem which ensures the existence of $n \times n$ matrix-valued function $\eta(\cdot,\mu):[-\tau,0]\rightarrow\mathbb{R}^{n^2}$, such that all the components of $\eta$ have bounded variations and for all $\phi \in C[-\tau,0]$
\begin{equation*}
  \mathcal{L}_\mu\phi = \int_{-\tau}^0d\eta(\theta,\mu)\phi(\theta),
\end{equation*}
where
\begin{equation*}
  d\eta(\theta,\mu) =\eta\big(\xi_y\delta(\theta+\tau_1)+\xi_z\delta(\theta+\tau_2)\big)d\theta,
\end{equation*}
and $\delta(\theta)$ is the Dirac delta function. Now we define the following operators for $\phi \in C[-\tau,0]$ 
\begin{equation}
\label{eq:Atheta}
  \mathcal{A}(\mu)\phi(\theta) =
  \begin{cases}
  \dfrac{d\phi(\theta)}{d\theta}, & \theta \in [-\tau,0)\\
  \int_{-\tau}^0d\eta(s,\mu)\phi(s), & \theta=0
  \end{cases}
\end{equation}
and
\begin{equation*}
  \mathcal{R}\phi(\theta) =
    \begin{cases}
  0, & \theta \in [-\tau,0)\\
  \mathcal{F}(\phi,\mu),& \theta=0.
  \end{cases}
\end{equation*}
Now the system (\ref{eq:auto_noneq}) becomes equivalent to (\ref{eq:auto_noneq_matrix}) as required. Let $q(\theta)$ be the eigenfunction for $\mathcal{A}(0)$ corresponding to $\lambda(0)$, namely
\begin{equation*}
\mathcal{A}(0)q(\theta) = i\omega_0q(\theta).
\end{equation*}
Now we define an adjoint operator $\mathcal{A}^*(0)$ as
\begin{equation*}
  \mathcal{A}^*(0)\alpha(s) =
    \begin{cases}
  -\dfrac{d\alpha(s)}{ds}, & s \in (0,\tau]\\
  \int_{-\tau}^0d\eta^T(t,0)\alpha(-t),& s=0
  \end{cases}
\end{equation*}
where $\eta^T$ denotes the transpose of $\eta.$
\newline Note  that the domains of $\mathcal{A}$ and $\mathcal{A}^*$ are $C^1[-\tau,0]$ and $C^1[0,\tau]$. As
\begin{equation*}
  \mathcal{A}q(\theta) = \lambda(0)q(\theta).
\end{equation*}
$\bar{\lambda}(0)$ is an eigenvalue for $\mathcal{A}^*$, and
\begin{equation*}
  A^*q^* = -i\omega_0q^*
\end{equation*}
for some nonzero vector $q^*$. For $\phi \in C[-\tau,0]$ and $\psi \in C[0,\tau]$, define a bilinear inner product
\begin{equation}\label{inner_pro}
  \varsigma \langle \psi,\phi\rangle = \bar{\psi}(0).\phi(0)-\int_{\theta=-\tau}^0\int_{\varsigma=0}^\theta\bar\psi^T(\varsigma-\theta)d\eta(\theta)\phi(\varsigma)d\varsigma.
\end{equation}
Then, $\langle\psi,A \phi\rangle = \langle A^*\psi,\phi\rangle $ for $\phi \in$ Dom$(\mathcal{A}),\psi \in$ Dom$(\mathcal{A}^*)$. Let $q(\theta) = e^{i\omega_0\theta}$ and $q^*(s) = De^{i\omega_0 s}$ be the eigenvectors for $ \mathcal{A}$ and $\mathcal{A}^*$ corresponding to the eigenvalues $+i\omega_0$ and $-i\omega_0$. Value of $D$ can be evaluated using (\ref{inner_pro}) and the relation $\langle q^*,q\rangle = 1$ as following
\begin{eqnarray}
& \hspace{-4mm} \langle q^*,q\rangle\, &\hspace{-4mm}=\bar{D}-\bar{D}\eta\int_{\theta = -\tau}^0\theta e^{i\omega_0\theta}\big(\xi_y\delta(\theta+\tau_1)+\xi_z\delta(\theta+\tau_2)\big)d\theta,\nonumber\\
& \hspace{-4mm}\Rightarrow 1&\hspace{-4mm}= \bar{D}+\bar{D}\eta\left(\tau_1\xi_ye^{-i\omega_0\tau_1}+\tau_2\xi_ze^{-i\omega_0\tau_2}\right),\nonumber\\
& \hspace{-4mm}\Rightarrow D&\hspace{-4mm}= \frac{1}{1+\eta\tau_1\xi_ye^{i\omega_0\tau_1}+\eta\tau_2\xi_ze^{i\omega_0\tau_2}}.\nonumber
\end{eqnarray}
Similarly, we can show that $\langle q^*,\bar{q}\rangle=0$. Now we define
\begin{equation}
\begin{aligned}
z(t) &= \langle q^*.u_t\rangle,  \\
w(t,\theta)&= u_t(\theta)-2Re\{z(t)q(\theta)\}.
\end{aligned}
\end{equation}
Then, on the center manifold $C_0$, $w(t,\theta)=w\big(z(t), \bar{z}(t),\theta\big)$, where
\begin{equation}\label{w_equ}
  w(z,\bar{z},\theta)=w_{20}(\theta)\frac{z^2}{2}+w_{11}(\theta)z\bar{z}+w_{02}(\theta)\frac{\bar{z}^2}{2}+\cdots.
\end{equation}
In effect, $z$ and $\bar{z}$ are local coordinates for manifold in $C$ in the directions
of $q^*$ and $\bar{q}^*$, respectively. The existence of the center
manifold $C_0$ enables us to reduce (\ref{eq:auto_noneq_matrix}) to an ordinary differential
equation for a single complex variable on $C_0$. At $\mu=0$, we have
\begin{eqnarray}
z'(t) &\hspace{-2mm}=&\hspace{-2mm} \langle q^*,\mathcal{A}y_t+\mathcal{R}u_t\rangle,\nonumber\\
&\hspace{-2mm}=&\hspace{-2mm} i\omega_0z(t)+\bar{q}^*(0).\mathcal{F}\big(w(z,\bar{z},\theta)+2Re\{z(t)q(\theta)\}\big),\nonumber\\
&\hspace{-2mm}=&\hspace{-2mm} i\omega_0z(t)+\bar{q}^*(0).\mathcal{F}_0(z,\bar{z}),\label{z_dot_prod}
\end{eqnarray}
which can be written as
\begin{equation}\label{abb_z}
  z'(t) = i\omega_0z(t)+g(z,\bar{z}).
\end{equation}
Now expanding the function $g(z,\bar{z})$ in powers of $z$ and $\bar{z}$ we get
\begin{eqnarray}
  g(z,\bar{z}) &=& \bar{q}^*(0).\mathcal{F}_0(z,\bar{z}),\nonumber\\
  &=& g_{20}\frac{z^2}{2}+g_{11}z\bar{z}+g_{02}\frac{\bar{z}^2}{2}+g_{21}\frac{z^2\bar{z}}{2}+\cdots.\nonumber
\end{eqnarray}
Following \citep{hassard1981}, we write
\begin{equation}\label{wdash}
  w' = u_t'-z'q-\bar{z}'\bar{q}.
\end{equation}
From (\ref{eq:auto_noneq_matrix}) and (\ref{abb_z}), we get
\begin{equation*}
  w' =
    \begin{cases}
  Aw-2Re\{\bar{q}^*(0).\mathcal{F}_0q(\theta)\}, & \theta \in [-\tau_2,0),\\
  Aw-2Re\{\bar{q}^*(0).\mathcal{F}_0q(0)\}+\mathcal{F}_0,& \theta=0,
  \end{cases}
\end{equation*}
which can be written as
\begin{equation}\label{eq:matrix_w}
  w' = Aw+H(z,\bar{z},\theta),
\end{equation}
using (\ref{abb_z}), where
\begin{equation}\label{H_equ}
  H(z,\bar{z},\theta)=H_{20}(\theta)\frac{z^2}{2}+H_{11}(\theta)z\bar{z}+H_{02}(\theta)\frac{\bar{z}^2}{2}+\cdots.
\end{equation}
Now, on the center manifold $C_0$, near the origin
\begin{equation*}
  w' = w_zz'+w_{\bar{z}}\bar{z}'.
 \end{equation*}
Use (\ref{w_equ}) and (\ref{abb_z}) to replace $w_z,\ z'$ and equating this with (\ref{eq:matrix_w}), we get
\begin{eqnarray}
  (2i\omega_0-A)w_{20}(\theta) &=& H_{20}(\theta),\label{eq1} \\
  -Aw_{11}(\theta) &=& H_{11}(\theta),\label{eq2}\\
  (2i\omega_0-A)w_{02}(\theta) &=& H_{02}(\theta).\label{eq2.1}
\end{eqnarray}
From (\ref{wdash}), we get
\begin{eqnarray}
  u_t(\theta) &=& w(z,\bar{z},\theta)+zq(\theta)+\bar{z}\bar{q}(\theta), \nonumber\\
   & =& w_{20}(\theta)\frac{z^2}{2}+w_{11}z\bar{z}+w_{02}(\theta)\frac{\bar{z}^2}{2}+ze^{i\omega_0\theta}+\bar{z}e^{-i\omega_0\theta}+\cdots,\nonumber
\end{eqnarray}
from which $u_t(0)$, $u_t(-\tau_1)$ and $u_t(-\tau_2)$ can be determined. 
As we only require the coefficients of $z^2,\ z\bar{z},\ \bar{z}^2$ and $z^2\bar{z}$, we have
\begin{eqnarray}
u_t(-\tau_1)\times u_t(-\tau_2) &\hspace{-1.5mm}=&\hspace{-1.5mm} \big(w(z,\bar{z},\tau_1)+ze^{-i\omega_0\tau_1}+\bar{z}e^{i\omega_0\tau_1}\big)\nonumber\\
&\hspace{-1.5mm}&\hspace{-1.5mm} \times \big(w(z,\bar{z},\tau_2)+ze^{-i\omega_0\tau_2}+\bar{z}e^{i\omega_0\tau_2}\big)\nonumber\\
&\hspace{-1.5mm}=&\hspace{-1.5mm} z^2e^{-i\omega_0(\tau_1+\tau_2)}+\bar{z}^2e^{i\omega_0(\tau_1+\tau_2)}\nonumber\\
&\hspace{-1.5mm}&\hspace{-1.5mm} +\ z\bar{z}(e^{-i\omega_0(\tau_1-\tau_2)}+e^{i\omega_0(\tau_1-\tau_2)})\nonumber\\   
&\hspace{-1.5mm}&\hspace{-1.5mm}+\ z^2\bar{z}\left(e^{-i\omega_0\tau_1}w_{11}(-\tau_2)\right.
\left.+\ e^{-i\omega_0\tau_2}w_{11}(-\tau_1)\right.\nonumber\\
&\hspace{-1.5mm}&\hspace{-1.5mm}\left.+\ e^{i\omega_0\tau_1}w_{20}(-\tau_2)/2\right)
\left.+\ e^{i\omega_0\tau_2}w_{20}(-\tau_1)/2\right)+\cdots.\nonumber\\
u_t(0) \times u_t(-\tau_1) &\hspace{-2mm}=&\hspace{-2mm} \big(w(z,\bar{z},0)+z+\bar{z}\big)\nonumber\\
\times u_t(-\tau_2)\quad&\hspace{-2mm}&\hspace{-2mm}\times \big(w(z,\bar{z},\tau_1)+ze^{-i\omega_0\tau_1}+\bar{z}e^{i\omega_0\tau_1}\big)\nonumber\\
&\hspace{-2mm}&\hspace{-2mm}\times(w(z,\bar{z},\tau_2)+ze^{-i\omega_0\tau_2}+\bar{z}e^{i\omega_0\tau_2})\nonumber\\
&\hspace{-2mm}=&\hspace{-2mm}z^2\bar{z}(e^{i\omega_0(\tau_1-\tau_2)}+e^{-i\omega_0(\tau_1-\tau_2)}+ e^{-i\omega_0(\tau_1+\tau_2)})+\cdots.\nonumber\\
u_t^3(-\tau_1) &\hspace{-2mm}=&\hspace{-2mm}  \big(w(z,\bar{z},\tau_1)+ze^{-i\omega_0\tau_1}+\bar{z}e^{i\omega_0\tau_1}\big)^3\nonumber\\
&\hspace{-2mm}=&\hspace{-2mm} 3z^2\bar{z}e^{-i\omega_0\tau_1}+\cdots;\nonumber.\\
u_t^2(-\tau_1)\times u_t(-\tau_2) &\hspace{-2mm}=&\hspace{-2mm} \big(w(z,\bar{z},\tau_1)+ze^{-i\omega_0\tau_1}+\bar{z}e^{i\omega_0\tau_1}\big)^2\nonumber\\
&\hspace{-2mm}&\hspace{-2mm}\times \big(w(z,\bar{z},\tau_2)+ze^{-i\omega_0\tau_2}+\bar{z}e^{i\omega_0\tau_2}\big)\nonumber\\
&\hspace{-2mm}=&\hspace{-2mm}z^2\bar{z}(e^{i\omega_0(-2\tau_1+\tau_2)}+2e^{-i\omega_0\tau_2})+\cdots.\nonumber\\
u_t(-\tau_1)\times u_t^2(-\tau_2) &\hspace{-2mm}=&\hspace{-2mm} \big(w(z,\bar{z},\tau_1)+ze^{-i\omega_0\tau_1}+\bar{z}e^{i\omega_0\tau_1}\big)\nonumber\\
&\hspace{-2mm}&\hspace{-2mm}\times \big(w(z,\bar{z},\tau_2)+ze^{-i\omega_0\tau_2}+\bar{z}e^{i\omega_0\tau_2}\big)^2\nonumber\\
&\hspace{-2mm}=&\hspace{-2mm}z^2\bar{z}(e^{i\omega_0(-2\tau_2+\tau_1)}+2e^{-i\omega_0\tau_1})+\cdots.\nonumber
\end{eqnarray}

Using the above equations, we can find the expressions for other quadratic and cubic terms of $u_t$.

Recall that
\begin{eqnarray}
   g(z,\bar{z}) &=& \bar{q}^*(0).\mathcal{F}_0(z,\bar{z}),\nonumber\\
   &=& g_{20}\frac{z^2}{2}+g_{11}z\bar{z}+g_{02}\frac{\bar{z}^2}{2}+g_{21}\frac{z^2\bar{z}}{2}+\cdots.\nonumber
\end{eqnarray}
Comparing the coefficients of $z^2,\ z\bar{z},\ \bar{z}^2$ and $z^2\bar{z}$, we get
\begin{eqnarray}
g_{20}=&&\hspace{-6mm}\bar{D}\eta[2\xi_{xy}e^{-i\omega_0\tau_1}+2\xi_{xz}e^{-i\omega_0\tau_2}\nonumber\\
&&\hspace{-6mm}+\ 2\xi_{yy}e^{-2i\omega_0\tau_1}+2\xi_{yz}e^{-i\omega_0(\tau_1+\tau_2)}+\ 2\xi_{zz}e^{-2i\omega_0\tau_2}]\label{g20}\nonumber\\
 g_{11}=&&\hspace{-6mm}\bar{D}\eta[\xi_{xy}(e^{-i\omega_0\tau_1}+e^{i\omega_0\tau_1})\nonumber\\
 &&\hspace{-6mm}+\ \xi_{xz}(e^{-i\omega_0\tau_2}+e^{i\omega_0\tau_2})+2\xi_{yy}\nonumber\\
 &&\hspace{-6mm}+\ \xi_{yz}(e^{-i\omega_0(\tau_1-\tau_2)}+e^{i\omega_0(\tau_1-\tau_2)})+ 2\xi_{zz}]\nonumber\\
 g_{02}=&&\hspace{-6mm}\bar{D}\eta[2\xi_{xy}e^{i\omega_0\tau_1}+2\xi_{xz}e^{i\omega_0\tau_2}\nonumber\\
 &&\hspace{-6mm}+\ 2\xi_{yy}e^{2i\omega_0\tau_1}+2\xi_{yz}e^{i\omega_0(\tau_1+\tau_2)}+\ 2\xi_{zz}e^{2i\omega_0\tau_2}]\nonumber.\\
g_{21}=&&\hspace{-6mm}\bar{D}\eta[\xi_{xy}\big(2w_{11}(0)e^{-i\omega_0\tau_1}+ w_{20}(0)e^{i\omega_0\tau_1}\nonumber\\
&&\hspace{-6mm}+2w_{11}(-\tau_1)+w_{20}(-\tau_1)\big)\nonumber\\
&&\hspace{-6mm}+\ \xi_{xz}\big(2w_{11}(0)e^{-i\omega_0\tau_2}+w_{20}(0)e^{i\omega_0\tau_2}\nonumber\\
&&\hspace{-6mm}+\ 2w_{11}(-\tau_2)+w_{20}(-\tau_2)\big)\nonumber\\
&&\hspace{-6mm}+\ \xi_{yy}\big(4w_{11}(-\tau_1)e^{-i\omega_0\tau_1}+2w_{20}(-\tau_1)e^{i\omega_0\tau_1}\big)\nonumber\\
&&\hspace{-6mm}+\ \xi_{yz}\big(2w_{11}(-\tau_1)e^{-i\omega_0\tau_2}+w_{20}(-\tau_1)e^{i\omega_0\tau_2}\nonumber\\
&&\hspace{-6mm}+\ 2w_{11}(-\tau_2)e^{-i\omega_0\tau_1}+w_{20}(-\tau_2)e^{i\omega_0\tau_1}\big)\nonumber\\
&&\hspace{-6mm}+\ \xi_{zz}\big(4w_{11}(-\tau_2)e^{-i\omega_0\tau_2}+ 2w_{20}(-\tau_2)e^{i\omega_0\tau_2}\big)\nonumber\\
&&\hspace{-6mm}+\ \xi_{xyy}(2e^{-2i\omega_0\tau_1}+4)+\ \xi_{xzz}(2e^{-2i\omega_0\tau_2}+4) \nonumber\\
&&\hspace{-6mm}+\ \xi_{yyz}(2e^{i\omega_0(-2\tau_1+\tau_2)}+4e^{-i\omega_0\tau_2})\nonumber\\ 
&&\hspace{-6mm}+\ \xi_{yzz}(2e^{i\omega_0(-2\tau_2+\tau_1)}+4e^{-i\omega_0\tau_1})\nonumber\\ 
&&\hspace{-6mm}+\ \xi_{xyz}(2e^{i\omega_0(\tau_1-\tau_2)}+2e^{-i\omega_0(\tau_1-\tau_2)}\nonumber\\ 
&&\hspace{-6mm}+\  2e^{-i\omega_0(\tau_1+\tau_2)})+6\xi_{yyy}e^{-i\omega_0\tau_1}+\ 6\xi_{zzz}e^{-i\omega_0\tau_2}].\label{g21}
\end{eqnarray}

For $\theta \in [-\tau,0)$, we have
\begin{eqnarray*}
H(z,\bar{z},\theta) &&\hspace{-6mm}= -2Re\{\bar{q}^*(0).\mathcal{F}_0q(\theta)\},\\
&&\hspace{-6mm}=-g(z,\bar{z})q(\theta) -\bar{g}(z,\bar{z})\bar{q}(\theta),\\
&&\hspace{-6mm}=-\left(g_{20}\frac{z^2}{2}+g_{11}z\bar{z} + g_{02}\frac{\bar{z}^2}{2}+\cdots\right)q(\theta)\\
 &&\hspace{-2mm}-\left(\bar{g}_{20}\frac{\bar{z}^2}{2}+\bar{g}_{11}z\bar{z} + \bar{g}_{02}\frac{z^2}{2}+\cdots\right)\bar{q}(\theta).
\end{eqnarray*}
Now using (\ref{H_equ}), we obtain
\begin{eqnarray*}
  H_{20}(\theta) &&\hspace{-6mm}= -g_{20}q(\theta)-\bar{g}_{20}\bar{q}\theta,   \\
  H_{11}(\theta) &&\hspace{-6mm}= -g_{11}q(\theta)-\bar{g}_{11}\bar{q}\theta.
\end{eqnarray*}
From (\ref{eq:Atheta}), (\ref{eq1}) and (\ref{eq2}), we derive the following
\begin{eqnarray*}
  w'_{20}(\theta) =&&\hspace{-6mm} 2i\omega_0w_{20}(\theta)+g_{20}q(\theta)+\bar{g}_{02}\bar{q}(\theta) \label{eq4},\\
  w'_{11}(\theta) =&&\hspace{-6mm} g_{11}q(\theta)+\bar{g}_{11}\bar{q}(\theta).\label{eq5}
\end{eqnarray*}
Solving the above differential equations yields
\begin{eqnarray}
w_{20}(\theta) &\hspace{-3mm}=&\hspace{-3mm} -\frac{g_{20}}{i\omega_0}q(0)e^{i\omega_0\theta} -\frac{\bar{g}_{02}}{3i\omega_0}\bar{q}(0)e^{-i\omega_0\theta}+Ee^{2i\omega_0\theta},\nonumber\\&&\label{w20}\\
w_{11}(\theta) &\hspace{-3mm}=&\hspace{-3mm} \frac{g_{11}}{i\omega_0}q(0)e^{i\omega_0\theta} -\frac{\bar{g}_{11}}{i\omega_0}\bar{q}(0)e^{-i\omega_0\theta}+F,\label{w11}
\end{eqnarray}
for some $E$ and $F$. For $\theta=0$, we get
\begin{eqnarray}
  H(z,\bar{z},0)=&&\hspace{-6mm} -2Re(\bar{q}^*.\mathcal{F}_0q(0))+\mathcal{F}_0,\nonumber\\
  H_{20}(0) =&&\hspace{-6mm} -g_{20}q(0) -\bar{g}_{20}\bar{q}(0)+\frac{g_{20}}{\bar{D}},\label{H_20}\\
  H_{11}(0) =&&\hspace{-6mm} -g_{11}q(0) -\bar{g}_{11}\bar{q}(0)+\frac{g_{11}}{\bar{D}}.\label{H_11}
\end{eqnarray}
Using (\ref{eq:Atheta}), (\ref{eq1}) and (\ref{eq2}), we get
\begin{eqnarray}
\eta\xi_yw_{20}(-\tau_1)+\eta\xi_zw_{20}(-\tau_2)
  &&=g_{20}q(0)+\bar{g}_{02}\bar{q}(0)-2\eta\big[\xi_{xy}e^{-i\omega_0\tau_1}\nonumber\\
  -2i\omega_0w_{20}(0)&&\ +\xi_{xz}e^{-i\omega_0\tau_2}+\ \xi_{yy}e^{-2i\omega_0\tau_1}+\xi_{yz}e^{-i\omega_0(\tau_1+\tau_2)}\nonumber\\&&\ +\ \xi_{zz}e^{-i\omega_0\tau_2}\big]\label{eq8} \\
\eta\xi_yw_{11}(-\tau_1)+\eta\xi_zw_{11}(-\tau_2)
  &&=  g_{11}q(0)+\bar{g}_{11}\bar{q}(0)-\eta\big[2\xi_{xy}e^{-i\omega_0\tau_1}\nonumber\\
  &&\ +\ 2\xi_{xz}e^{-i\omega_0\tau_2}+2\xi_{yy}e^{-2i\omega_0\tau_1}\nonumber\\
  &&\ +\ 2\xi_{yz}e^{-i\omega_0(\tau_1+\tau_2)}+2\xi_{zz}e^{-i\omega_0\tau_2}\big]\label{eq9}.
  \end{eqnarray}

Evaluate $w_{11}(0)$, $w_{20}(0)$, $w_{11}(-\tau_1)$, $w_{20}(-\tau_1)$,  $w_{11}(-\tau_2)$ and $w_{20}(-\tau_2)$ using (\ref{w20}) and (\ref{w11}), and substituting in (\ref{eq8}) and (\ref{eq9}), we get $E$ and $F$ as\\
\begin{eqnarray*}
  && E = \frac{-g_{20}}{\bar{D}(\eta\xi_ye^{-2i\omega_0\tau_1}+\eta\xi_ze^{-2i\omega_0\tau_2}-2i\omega_0)}, \\
&& F = \frac{-g_{11}}{\bar{D}\eta(\xi_y+\xi_z)}.
\end{eqnarray*}
Thus the local Hopf bifurcation analysis can now be performed using \citep{hassard1981}.
The quantities required to study the nature of the Hopf bifurcation are as follows\\
\begin{eqnarray}
\mu_2 &&\hspace{-6mm}= \dfrac{-\mathbf{Re}\big(c_1(0)\big)}{\alpha'(0)},\quad \beta_2= 2\mathbf{Re}\big(c_1(0)\big),\nonumber
\end{eqnarray}
where

\begin{align*}
\alpha'(0) &=  \mathbf{Re}\left(\dfrac{d\lambda}{d\eta}\right)_{\eta=\eta_c} > 0,\\
c_1(0) &= \dfrac{i}{2\omega_0}\left(g_{20}g_{11}-2|g_{11}|^2-\dfrac{1}{3}|g_{02}|^2\right)+\dfrac{g_{21}}{2}.\nonumber\label{c10}\\
\end{align*}
The direction and stability of Hopf bifurcation can be determined by the sign of $\mu_2$ and  $\beta_2$ respectively.
If $\mu_2 > 0\ (\mu_2 < 0)$ then the Hopf bifurcation is \emph{super-critical (sub-critical)}. Similarly, the bifurcating solutions are \emph{asymptotically orbitally stable (unstable)} if $\beta_2 < 0\ (\beta_2>0)$.\\


In \citep{raina2005}, the Hopf bifurcation properties a non-linear equation with a single discrete delay was studied in some detail. The analysis allowed us to ascertain that some non-linear terms always produced a Hopf bifurcation of certain type. This enabled us to identify the impact of some non-linear terms on the nature of Hopf bifurcation.

This paper considers a non-linear equation with two discrete delays. So leading from the previous work \citep{raina2005}, it might be natural to ask if we may develop a similar understanding for the two delay variant. We may ask: If we choose a particular non-linear equation, can we change the type of Hopf bifurcation by simply changing the delays? Let us consider some examples. \\ \\
\textbf{Example A.1} Consider the following non-linear delay equation\\
\begin{equation}
 \frac{d}{dt}u(t) = \eta \big( -a u(t-\tau_1) - a u(t-\tau_2) + \gamma u^2(t-\tau_2) \big), \label{eq:appeqn1}
\end{equation}
where $a$, $\gamma$, $\eta$, $\tau_1$, $\tau_2 > 0$. Using $\eta$ as a bifurcation parameter that drives the above equation just beyond the locally stable regime. The linearized equation associated with \eqref{eq:appeqn1} is
\begin{equation}
 \frac{d}{dt}u(t) = \eta ( -a u(t-\tau_1) -a u(t-\tau_2))\label{eq:appeqn1lin}
\end{equation}
with the following characteristic equation
\begin{equation}
 \lambda + a \eta e^{-\lambda \tau_1} + a  \eta e^{-\lambda \tau_2} = 0.
\end{equation}
A necessary and sufficient condition for the stability of the equation \eqref{eq:appeqn1lin} is  \citep{stepan1989}
\begin{equation}
\label{eq:appns_pf} 
\eta a (\tau_1+\tau_2)\cos\left(\dfrac{\pi(\tau_1-\tau_2)}{2(\tau_1+\tau_2)}\right)<\dfrac{\pi}{2}, 
\end{equation}
and the Hopf bifurcation occurs at 
\begin{equation}
\label{eq:appkappa_cval} 
\eta_c=\dfrac{\pi/2}{a(\tau_1+\tau_2)\cos\left(\dfrac{\pi(\tau_1-\tau_2)}{2(\tau_1+\tau_2)}\right)}
\end{equation}
with period $2(\tau_1+\tau_2)$, where $\eta_c$ is the critical value of $\eta$ which induces a Hopf bifurcation.\\

Using the framework outlined in this paper, we calculate the value of $\mu_2$ for the system \eqref{eq:appeqn1} as
\begin{equation}
 \mu_2 = \frac{2 \eta \gamma^2}{\pi a^2 \sin(\theta)\big( 4 \cos^2(2\theta) + 16 \sin^2(\theta) \big)} \times \tilde{s}(\theta),
\end{equation}\
where
\begin{eqnarray}
 \tilde{s}(\theta) &&=- \Bigg[ \Big( 2+ \cos(\theta)-2\cos(2\theta) + 2\cos(3\theta)\Big)\\ \nonumber
 &&\hspace{4mm}  \times \Big( \sin(\theta) +2\sin(2\theta) + 2\sin(3\theta) \Big)\\ \nonumber
&&\hspace{4mm} + \Big(2\cos(2\theta) \big(1 + \cot(\theta)(\pi/2 - \theta)  \big) - 2\pi \sin(\theta)\Big)\\ \nonumber
 &&\hspace{4mm}  \times \Big( 4\sin(\theta)\big( 1 + \cot(\theta)(\pi/2 - \theta)) \Big) \Bigg],\\
 \theta  &&= \omega_0\tau_2 = \dfrac{\pi\tau_2}{\left(\tau_1+\tau_2\right)} \in (0,\pi) \quad \forall \ \tau_1, \tau_2 > 0.\label{eq:app_eg1}\\\nonumber
\end{eqnarray}
As $\eta$, $a$ and $\gamma >0$, we get
\begin{align*}
\mathop{\mathrm{sign}}\Big(\mu_2\Big) =& \mathop{\mathrm{sign}}\Big(\tilde{s}(\theta)\Big).
\end{align*}

\begin{figure}[hbtp!]
\center
\psfrag{x}{$\theta$}
\psfrag{y}[c]{$\tilde{s}(\theta)$}
\psfrag{0.000}{\small{\hspace{2.5mm}$0$}}
\psfrag{0.785}{\small{\hspace{1mm}$\pi/4$}}
\psfrag{1.570}{\small{\hspace{1mm}$\pi/2$}}
\psfrag{2.356}{\small{$3\pi/4$}}
\psfrag{3.141}{\small{\hspace{2.5mm}$\pi$}}
\psfrag{0.5}{\small{\hspace{0mm}$0.5$}}
\psfrag{1.0}{\small{\hspace{0mm}$1.0$}}
\psfrag{1.5}{\small{\hspace{0mm}$1.5$}}
\psfrag{beta}{\small{$\beta_2$}}
\psfrag{myu2}{\small{$\mu_2$}}
\psfrag{105}{{\small{$\times 10^{-5}$}}}
\psfrag{107}{{\small{$\times 10^{-7}$}}}
\includegraphics[trim=0cm 0cm 0cm 1cm, clip=true,width=3.25in,height=2.25in]{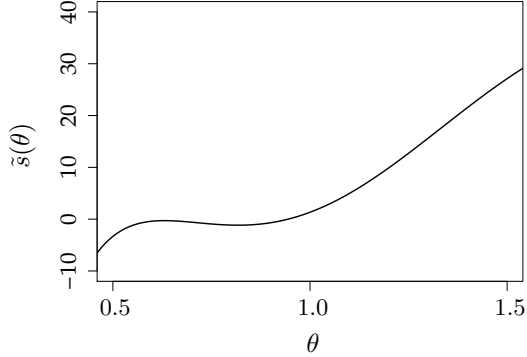}
\vspace{-2mm}
\caption{The plot of $\tilde{s}(\theta)$ as $\theta$ varies, where $\mathop{\mathrm{sign}}(\mu_2) = \mathop{\mathrm{sign}}\big(\tilde{s}(\theta)\big)$ and $\theta = (\pi\tau_2)/\left(\tau_1+\tau_2\right)$. Observe that $\tilde{s}(\theta)$ changes the sign from negative to positive at $\theta=\theta_h \approx 0.95$. Hence, the criticality of the Hopf bifurcation changes from sub-critical to super-critical at $\theta=0.95$.}
\label{fig:apprealc10}
\end{figure}

From Fig. \ref{fig:apprealc10}, we can observe that the sign of $\tilde{s}(\theta)$ changes as $\theta$ increases. We evaluate that the sign of $\tilde{s}(\theta)$ changes from negative to positive at $\theta = \theta_h \approx 0.95$. Thus, the Hopf bifurcation is sub-critical ($\mu_2<0$) for $\theta<\theta_h$, and is super-critical for $\theta>\theta_h$. Substituting $\theta = \theta_h$ in \eqref{eq:app_eg1}, we get the ratio of the delays at which the criticality of Hopf changes as  ($\tau_1/\tau_2) = 2.306$. Therefore, the system undergoes a super-critical Hopf for ($\tau_1/\tau_2) < 2.306$, and a sub-critical Hopf for ($\tau_1/\tau_2) > 2.306$.\\ 

We now summarize the inferences from the results of Hopf bifurcation analysis for the system \eqref{eq:appeqn1} as follows.
\begin{itemize}
   \item The values of the parameters $a$ and $\gamma$ do not affect the sign of $\mu_2$. Hence, the type of Hopf bifurcation is independent of these parameters.
  \item The nature of Hopf bifurcation depends only on the ratio of the delays. For ($\tau_1/\tau_2) < 2.303$, we have $\mu_2>0$ and hence the bifurcation is super-critical Hopf. Whereas, the criticality of the bifurcation changes to sub-critical Hopf for ($\tau_1/\tau_2) > 2.303$.
\end{itemize}

%
%
%
%
%
%

To validate this, we consider some special cases where we choose specific values for the delays.\\

\textit{Case} (1): Let $\tau_1=3$ and $\tau_2=1$.\\
The Hopf condition is: $\eta_c = \pi \sqrt{2}/8$, where $\eta_c$ denotes the critical value of $\eta$ which induces a Hopf bifurcation. For these values, we get 
\begin{equation*}
 \alpha^{'}(0) = \mathbf{Re}\left(\frac{d\lambda}{d\eta}\right)_{\eta=\eta_c} = \frac{8\pi\sqrt{2}}{16+8\pi+5\pi^2} > 0,
\end{equation*}
The Hopf bifurcation is super-critical if $\mu_2>0$ and sub-critical if $\mu_2<0$, where 
\begin{equation}
 \mu_2 =  - 0.0499(\gamma/a)^2 < 0,
\end{equation}
which indicates that we will get a sub-critical Hopf bifurcation.\\

\textit{Case} (2): Let $\tau_1=2$ and $\tau_2=1$.\\ 
The Hopf condition is: $\eta_c = \pi\sqrt{3}/9$, where $\eta=\eta_c+\mu$ with a Hopf bifurcation taking place at $\mu=0$. We get 
\begin{equation*}
 \alpha^{'}(0) = \mathbf{Re}\left(\frac{d\lambda}{d\eta}\right)_{\eta=\eta_c} = \frac{27\pi\sqrt{3}}{54+6\pi\sqrt{3}+14\pi^2} > 0,
\end{equation*}
\begin{equation}
 \omega_0 = \frac{\pi}{3}, \quad  \bar{D} = \frac{3(18+\pi\sqrt{3})- i 27\pi}{54+6\pi\sqrt{3}+14\pi^2}.
\end{equation}
The Hopf bifurcation is super-critical if $\mu_2>0$ and sub-critical if $\mu_2<0$, where 
\begin{equation}
 \mu_2 =  0.0846(\gamma/a)^2 > 0,
\end{equation}
which implies that the system undergoes a super-critical Hopf bifurcation.\\ \\
Remark: In a scalar non-linear equation with two discrete delays, both the delays play an important role in determining the type of the Hopf bifurcation. As we have just witnessed that even with the same non-linear term, by simply changing the values of the delay, we can change the type of the Hopf bifurcation.
\balance

\end{document}